# A long-lasting guided bone regeneration membrane from sequentially functionalised photoactive atelocollagen

He Liang,[a,b,1] Jie Yin,[a,b,2] Kenny Man,[a,3] Xuebin B. Yang,[a] Elena Calciolari,[c] Nikolaos Donos,[c] Stephen J. Russell,[b] David J. Wood,[b] Giuseppe Tronci[a,b*]

[a] Biomaterials and Tissue Engineering Research Group, School of Dentistry, St. James's University Hospital, University of Leeds, United Kingdom

[b] Clothworkers' Centre for Textile Materials Innovation for Healthcare, School of Design, University of Leeds, United Kingdom

[c] Centre for Oral Immunobiology & Regenerative Medicine & Centre for Oral Clinical Research, Barts and The London School of Medicine & Dentistry, Institute of Dentistry, Queen Mary University of London, United Kingdom

## Abstract

The fast degradation of collagen-based membranes in the biological environment remains a critical challenge, resulting in underperforming Guided Bone Regeneration (GBR) therapy leading to compromised clinical results. Photoactive atelocollagen (AC) systems functionalised with ethylenically unsaturated monomers, such as 4-vinylbenzyl chloride (4VBC), have been shown to generate mechanically competent materials for wound healing, inflammation control and drug delivery, whereby control of the molecular architecture of the AC network is key. Building on this platform, the sequential functionalisation with 4VBC and methacrylic anhydride (MA) was hypothesised to generate UV-cured AC hydrogels with reduced swelling ratio, increased proteolytic stability and barrier functionality for GBR therapy. The <u>s</u>equentially functionalised <u>a</u>telocollagen <u>p</u>recursor (SAP) was characterised via TNBS and ninhydrin colourimetric assays, circular dichroism and UV-curing rheometry, which confirmed nearly complete consumption of collagen's primary amino groups, preserved triple helices and fast (< 180 s) gelation kinetics, respectively. Hydrogel's swelling ratio and compression modulus were adjusted depending on the aqueous environment used for UV-curing, whilst the sequential functionalisation of AC successfully generated hydrogels with superior proteolytic stability *in vitro* compared to both 4VBC-functionalised control and the commercial dental membrane Bio-Gide®. These *in vitro* results were confirmed *in vivo* via both subcutaneous implantation and a proof-of-concept study in a GBR calvarial model, indicating integrity of the hydrogel and barrier defect, as well as tissue formation following 1-month implantation in rats.

---

* Corresponding author. Level 7 Wellcome Trust Brenner Building, St. James's University Hospital, Leeds, LS9 7TF, United Kingdom.
[1] Present address: 4D Biomaterials, 97 Vincent Drive, Edgbaston, Birmingham, B15 2SQ, United Kingdom
[2] Present address: Henkel (China) Investment Co., Ltd, Shanghai, China
[3] Present address: School of Chemical Engineering, University of Birmingham, United Kingdom
Email addresses: h.liang@4dbiomaterials.co.uk (H. Liang), jie.yin@henkel.com (J. Yin), k.l.man@bham.ac.uk (K. Man), x.b.yang@leeds.ac.uk (X. Yang), e.calciolari@qmul.ac.uk (E. Calciolari), n.donos@qmul.ac.uk (N. Donos), s.j.russell@leeds.ac.uk (S.J. Russell), d.j.wood@leeds.ac.uk (D.J. Wood), g.tronci@leeds.ac.uk (G. Tronci).

## 1. Introduction

Guided Bone Regeneration (GBR) therapy has gained increasing attention, due to the growing trends in bone disorders associated with ageing, congenital diseases, trauma, and cancer, and the avoidance of tissue morbidity commonly observed with autologous bone grafting [1,2]. GBR therapy has proven successful in supporting the repair of alveolar bone critical-sized defects in the oral cavity [3], whilst new GBR strategies integrating membranes and soluble factors have also been explored for orthopaedic reconstruction [4,5]. An essential component of GBR therapy in dentistry is the placement of a physical barrier between the connective tissue and the critical-sized alveolar bone defect to allow repopulation of region-specific osteogenic cells in the bone defect [6,7]. The GBR membrane acts as a barrier against soft tissue cells present in the connective tissue, aiming to promote the selective proliferation of osteogenic cells. Together with the GBR membrane, a graft material may also be applied to the critical-sized bone defect *in vivo* to provide an osteoconductive environment for bone regeneration. In addition, the graft material may also be applied to ensure membrane conformation to, and preservation of, the critical-sized defect, minimising risks of membrane collapse into the defect [8,9].

Resorption of the GBR membrane *in vivo* is a key aim in promoting full bone regeneration with a single-step clinical procedure, minimising the infection risks associated with a second surgery and membrane removal [10]. Although an ideal degradation profile has yet to be established, it is well accepted that the membrane must retain its barrier function for a certain amount of time following implantation *in vivo* to achieve a predictable regenerative outcome [11,12]. Considering that completed bone regeneration can take up to six months depending on the defect size [13], it can be estimated that the GBR membrane should display material integrity in the range from four weeks, when the bone remodelling phase starts, up to several months, to preserve the volume of the bone tissue defect, support selective homing of osteogenic cells and avoid soft tissue infiltration. Overall, the membrane should display (i) elasticity, to enable clinical handling and implantation *in vivo* with no mechanical damage; (ii) space-maintaining capability, to ensure conformation with, and protection of, the bone defect; (iii) tissue occlusivity, to minimise risks of soft tissue infiltration following implantation and during degradation; (iv) safe degradability to minimise risks of eliciting toxic response during membrane degradation [14]. Although GBR membranes have been used widely to treat bone defects in relation to severe periodontitis, peri-implantitis, and large bone loss, long term stability *in vivo* (> 1 month) is still a challenge for most clinically approved resorbable GBR products [14,15,16,17], resulting in soft tissue ingrowth and underperforming clinical outcome [18].

Multiple natural polymers, e.g. bacterial cellulose [2], chitin [19], hyaluronic acid [20], poly($\gamma$-glutamic acid) [21], and silk [22], have recently been proposed as protease-insensitive substrates for the design of new resorbable GBR membranes with prolonged stability in the physiological environment. Customisation of multiphase formulations into bespoke fibrous configurations has proven promising, aiming to induce membrane tissue adhesion, trigger an osteogenic environment *in vitro* and minimise risks of surgery-induced infection [6,22,23,24,25]. These approaches have subsequently been translated to synthetic polymers aiming to overcome the potential toxicity of degradation products [17], and to accomplish mechanical competence and space maintenance capability, together with osteogenic and barrier functionalities *in vitro* [26,27,28]. While these research efforts have shown promise at the preclinical testing stage, clinically approved resorbable membranes are still mainly made of type I collagen [7,14,18]. This is largely attributed to the fact that type I collagen is the main protein component of the extracellular matrix (ECM) of bone, whereby the hierarchical structure and natural crosslinks of collagen regulate bone function and remodelling *in vivo*. Extraction of collagen *ex vivo* is generally achieved via incubation in acidic or enzymatic media, which results in the breakdown of natural crosslinks of collagen [29]. The resulting water-soluble material therefore presents minimal applicability in the physiological environment, so that synthetic routes aiming to restore covalent crosslinks are key [30].

Chemical crosslinking of type I collagen *ex vivo* has been widely adopted to equip resulting materials with water insolubility and mechanical competence in water [31], although the limited chemical accessibility of collagen triple helices raises challenges in the control and adjustment of macroscopic properties [32]. Recent synthetic approaches to accomplish controlled elasticity and degradation include crosslinked fibrillated type I collagen systems [33], UV-cured methacrylated collagen networks [34,35], collagen crosslinking with macromolecular segments [36,37,38,39], ionic co-networks [40], and nanostructured collagen-polyester composite [41]. While single-phase collagen materials proved to display at least 30 wt.% mass loss following 48 hours in collagenase medium [33,34], the fabrication of composite materials enabled structural customisation in the form of e.g. fibrous membrane [15] and double-layer sponge [42], whereby a mass loss of up to 30 wt.% was measured following 30 days *in vitro* [41]. These reports therefore indicate that current synthetic routes still only partially enable the fabrication of single-phase collagen-based membranes with GBR-compatible resorption rate (> 1 month), whilst the use of secondary polymeric phase, e.g. aliphatic polyesters, is critical to ensure long-lasting membrane stability *in vivo*, with the consequent safety concerns of degradation products.

In an effort to address this challenge, we have recently developed a family of telopeptide-free collagen (i.e. AC) systems functionalised with ethylenically unsaturated monomers, e.g.

4-vinylbenzyl chloride, that successfully generate hydrogel networks under light exposure with no addition of secondary crosslinker [ 43 , 44 ]. This strategy directly enabled customisation of properties according to the type of monomer that was chemically coupled to the collagen backbone [43, 45 , 46 ]. Additional biofunctionalities could be successfully introduced in a drug-free fashion by exploiting monomer-induced secondary interactions. For example, these molecular effects could generate drug-free inhibition of matrix metalloproteinases (MMPs) [44,47,48], ultimately delaying the enzymatic cleavage of the collagen network *in vivo* for durable material stability. The aim of this work was therefore to realise a single-phase and long-lasting AC-based hydrogel network with GBR applicability by leveraging the knowledge gained with the monomer-induced customisation of AC photonetworks. Departing from current design approaches of three-dimensional fibrous assemblies [16], hybrid synthetic/natural polymer formulations [41] and multi-layer composite structures [42], we hypothesised that building a covalent network of collagen triple helices with controlled crosslink density, capable of mediating drug-free MMP inhibition could enable the fabrication of a hydrogel-based GBR membrane with competitive elasticity, long-lasting stability *in vivo* and soft tissue barrier functionality. Our design strategy to deliver this was centred on the sequential functionalisation of AC triple helices with both 4VBC, as the MMP-inhibiting monomer, and MA, as the highly reactive monomer, and to investigate the effect of the UV-curing solvent as an additional experimental space to adjust hydrogel compressibility, proteolytic degradability and swelling properties. The thorough physicochemical characterisation was pursued (i) to identify specific atelocollagen formulations generating GBR-compliant membrane microstructure and macroscopic properties (i.e. long-lasting stability, relatively low swelling and mechanical competence in biological environments); (ii) demonstrate *in vitro* the superior prototype proteolytic stability over Bio-Gide® as one of the clinical gold standard membranes; and (iii) corroborate *in vivo* the direct effect played by the prototype properties on the GBR performance, aiming to de-risk future product development.

## 2. Materials and Methods

### 2.1 Materials

Pepsin-extracted, medical-grade type I AC (6 mg·ml$^{-1}$ in 10 mM hydrochloric acid) was purchased from Collagen Solutions PLC (Glasgow, UK). 4VBC, MA and triethylamine (TEA) were purchased from Sigma-Aldrich (UK). 2-Hydroxy-1-[4-(2-hydroxyethoxy) phenyl]-2-methylpropan-1-one (I2959) was purchased from Fluorochem Limited (Glossop, UK). Ninhydrin 99% was ordered from Alfa-Aesar (Massachusetts, USA). Absolute ethanol and diethyl ether were purchased from VWR internationals. All other chemicals were purchased from Sigma-Aldrich unless specified.

### 2.2 Sequential functionalisation of AC

A two-step functionalisation route was developed to accomplish covalent coupling of both 4VBC and MA residues to the AC backbone. In the first step, 4VBC-functionalised AC was prepared as previously reported [43,44]. The solution of AC was diluted to a concentration of 3 mg·ml$^{-1}$ by addition of 10 mM HCl and subsequently neutralised to pH 7.5 by the addition of 0.1 M NaOH and 10 mM hydrochloric acid (HCl). Polysorbate 20 (PS-20) was introduced at a concentration of 1 wt.% concentration (with respect to the weight of the diluted AC solution), prior to the addition of 4VBC and TEA at a fixed molar ratio of 4VBC (and TEA) with respect to AC's primary amino groups ([4VBC]·[Lys]$^{-1}$ = 25; [4VBC] = [TEA]). After 24-hour reaction, the mixture was precipitated in 10-volume excess of absolute ethanol for at least 8 hours, recovered by centrifugation and air-dried. The freshly synthesised 4VBC-functionalised AC was subsequently solubilised in 10 mM HCl (3 mg·ml$^{-1}$) under stirring at room temperature, and the solution pH adjusted to 7.5. MA and TEA were then added at a molar ratio of 25 ([MA]·[Lys]$^{-1}$ = 25; [MA] = [TEA]). After a 24-hour reaction, the reacting mixture was precipitated in 10-volume excess of absolute ethanol for 8 hours, recovered by centrifugation and air-dried. Samples of the sequentially functionalised AC and 4VBC-functionalised control are coded as *4VBC-MA* and *4VBC*, respectively.

## 2.3 (2,4,6)-Trinitrobenzenesulfonic acid (TNBS) and Ninhydrin assays

Both TNBS and Ninhydrin assays were used to measure the consumption of free AC's amino groups and respective degree of AC functionalisation, as described previously [41,42,45]. For the TNBS assay, the molar content of primary free amino groups and degree of functionalisation (*F*) were calculated via Equation 1 and Equation 2, respectively:

$$\frac{mol(Lys)}{g(AC)} = \frac{2 \times Abs(346nm) \times 0.02}{1.46 \cdot 10^4 \times b \times x}$$ (Equation 1)

$$F = 100 - \frac{mol(Lys)_{funct.}}{mol(Lys)_{AC}} \times 100$$ (Equation 2)

where *Abs (346nm)* is the absorbance value recorded at 346 nm; *2* is the dilution factor; *0.02* is the volume of the sample solution (in litres); *1.46·10$^4$* is the molar absorption coefficient for 2,4,6-trinitrophenyl lysine (in M$^{-1}$ cm$^{-1}$); *b* is the cell path length (1 cm); *x* is the dry sample weight; and *mol(Lys)$_{AC}$* and *mol(Lys)$_{funct}$* represent the total molar content of free amino groups in native and functionalised AC, respectively.

Confirmation of collagen functionalisation was obtained via Ninhydrin assay. 10 mg of the dry sample was mixed with 4 ml of distilled water and 1 ml of 8 wt.% Ninhydrin solution in acetone. The mixture was reacted at 100 °C for 15 mins, followed by cooling in ice and the addition of 1 ml of 50 vol.% ethanol in distilled water. The molar content of amino groups was measured by reading the absorbance at 570 nm against the blank. A standard calibration curve was prepared with collagen samples of known mass.

Three replicates were used for each assay, and data presented as mean ± SD.

## 2.4 Circular dichroism

The circular dichroism (CD) spectra of native and functionalised AC samples were acquired using a Chirascan CD spectrometer (Applied Photophysics Ltd) using 0.2 mg·ml$^{-1}$ solutions in 17.4 mM acetic acid (AcOH) solution [28]. Sample solutions (n = 2) were collected in quartz cells of 1.0 mm path length, whereby CD spectra were obtained with 2.0 nm band width and 20 nm·min$^{-1}$ scanning speed. A spectrum of the 17.4 mM acetic acid control solution was subtracted from each sample spectrum. The mean residue ellipticity ($\theta_{mrw,\lambda}$) was calculated according to Equation 3:

$$\theta_{mrw,\lambda} = \frac{MRW \times \theta_\lambda}{10 \times d \times c} \quad \text{(Equation 3)}$$

where *MRW* is the mean residue weight and equals to 91 g·mol$^{-1}$ for amino acids [47]; $\theta_\lambda$ is the observed ellipticity (degrees) at wavelength $\lambda$, *d* is the path length (1.0 mm), and *c* is the concentration (0.2 mg·ml$^{-1}$).

## 2.5 Fabrication of the GBR membrane

Samples of SAP were dissolved (1.2 wt.%) in solutions of either 10 mM HCl, 17.4 mM AcOH or 10 mM phosphate-buffered saline (PBS), to investigate the effect of typical AC solvents on the macroscopic properties of resulting UV-cured AC hydrogels. I2959 was selected as a water-soluble photoinitiator and firstly dissolved (1 wt.%) in the aforementioned solvents via incubation at 60 ºC in the dark for three hours. Following equilibration to room temperature, the obtained I2959-supplemented solutions of either 10 mM HCl, 17.4 mM AcOH or 10 mM PBS were employed for the solubilisation of AC. The resulting AC solutions were centrifuged at 3000 rpm for 5 mins to remove any air bubbles and then cast onto a 24-well plate (Corning Costar), followed by curing under UV lamp (365 nm, 8 mW·cm$^{-2}$, Spectroline) for 30 mins on both top and bottom sides. At least three hydrogel replicates were prepared in each UV-curing solvent, whereby 0.6-0.8 g of hydrogel-forming solution were poured in each well prior to UV-curing. The resulting UV-cured samples of GBR membrane were carefully removed with a spatula and rinsed with distilled water (15 mins, x3). Dehydration in an ascending series of distilled water-ethanol mixtures (0, 20, 40, 60, 80, (×3) 100 wt.% EtOH) was carried out prior to air-drying. The above dehydration step was carried out to ensure gradual removal of water from the sample while preserving the sample geometry, as well as to remove any unreacted species. Control samples of 4VBC-functionalised AC were prepared as outlined above via solubilisation in either 10 mM HCl or 17.4 mM AcOH solutions supplemented with 1 wt.% I2959. Samples of UV-cured network are coded as either *4VBC-MA\** or *4VBC\**, where * indicates the crosslinked state.

## 2.6 Swelling, gel content and compression tests

Dry UV-cured samples of known mass ($m_d$= 10-20 mg, n=4) were individually incubated in distilled water (1.5 ml) at room temperature. Following 24-hour incubation, water-equilibrated samples were collected, paper blotted and weighed ($m_s$). The swelling ratio (*SR*) was calculated according to Equation 4:

$$SR = \frac{m_s - m_d}{m_d} \times 100 \qquad \textbf{(Equation 4)}$$

The gel content was measured to investigate the overall fraction of the covalent hydrogel network insoluble in 17.4 mM AcOH [43-47]. Dry collagen networks ($m_d$= 10-20 mg, n=4) were individually incubated in 2 ml of 17.4 mM AcOH for 24 hours, and subsequently air-dried and weighed ($m_1$). The gel content (*G*) was calculated according to Equation 5:

$$G = \frac{m_1}{m_d} \times 100 \qquad \textbf{(Equation 5)}$$

Water-equilibrated hydrogel discs (Ø: 14 mm; h: 5-6 mm, n=3) were compressed at room temperature with a compression rate of 3 mm·min$^{-1}$ (Instron ElectroPuls E3000) and a 250 N load cell. The dimensions of the hydrogel samples (Ø: 14 mm; h: 5-6 mm) were selected to minimise risks of sample tilting prior to or during the test, and to ensure uniaxial compression, in line with recent publications [36,43,49,66]. The compression modulus was quantified as the slope of the linear plot region at 25-45% compression in the stress-compression curves, at which range any potential risks of a platten effect are minimal.

Data of swelling ratio, gel content and compression modulus were presented as mean ± SD.

## 2.7 Scanning electron microscopy (SEM)

The morphology of UV-cured hydrogel surface and cross-section was explored via variable pressure SEM (Hitachi S-3400N VP, 60-70 Pa) combined with a Deben cool stage control (Model: LT3299).

## 2.8 Enzymatic degradability *in vitro*

Dry UV-cured samples (n=4) of known mass ($m_d$) were incubated for up to 14 days in 1 ml of 50 mM [tris(hydroxymethyl)-methyl-2-aminoethane sulfonate] (TES) buffer containing 0.36 mM calcium chloride at 37°C and either 1 or 2.5 collagen digestive units (CDU) per mg of the sample of collagenase type I from *Clostridium histolyticum* (125 CDU·mg$^{-1}$ solid). At the selected degradation time point (either day 4 or day 14), the samples were collected, rinsed with distilled water (15 min, x3), and dehydrated via an ascending series of water-ethanol mixtures (0, 20, 40, 60, 80, (×3) 100 vol.% EtOH) followed by air-drying and weighing ($m_t$). The relative mass ($\mu_{rel}$) of retrieved samples was calculated as below:

$$\mu_{rel} = \frac{m_t}{m_d} \times 100 \qquad \text{(Equation 6)}$$

**2.9 Gelation kinetics by UV-equipped rheometer**

The UV-induced gelation kinetics of the GBR membrane was measured by a modular compact rotational rheometer (MCR 302, Anton Paar, Austria) equipped with a UV curing module (Ominicure 1500, Excelitus Technologies). The measurements of storage modulus ($G'$) and loss modulus ($G''$) were accomplished during UV light exposure of the I2959-supplemented solution (~ 0.4 g) of functionalised AC, to characterise the gelation kinetics profile and confirm the synthesis of the UV-induced collagen network. The oscillatory shear was applied to a transparent glass parallel plate (Ø: 25mm) with a gap between the plates of 300 µm. UV light (365 nm, 8 mW·cm$^{-2}$) was initiated after 5 s of shear oscillation at 21˚C.

**2.10 Cytotoxicity tests**

L929 fibroblast cells were cultured in Dulbecco's modified Eagle's medium (DMEM), supplemented with 10% fetal bovine serum (FBS), 1% glutamine, and 2.5 mg·ml$^{-1}$ penicillin-streptomycin, in a humidified incubator (37°C, 5% $CO_2$). Cells were passaged every 3 days with 0.25% trypsin/0.02% EDTA. UV-cured samples were individually prepared on to a 24-well plate, extensively washed in distilled water and dehydrated in an ascending series of distilled water-ethanol mixtures [0, 20, 40, 60, 80, (×3) 100 vol.% EtOH]. This dehydration step was carried out aiming to accomplish gradual removal of water, preservation of the sample geometry, and removal of any unreacted species. Prior to cell seeding, samples were disinfected in a 70 vol.% ethanol solution under UV light and washed in PBS (10 min,

×3) under a cell culture hood. L929 cells (8×10$^3$ cells·ml$^{-1}$) were seeded on top of the hydrogels and left for 3 hours to allow time for cells to adhere, followed by the addition of 2 ml of cell culture medium and overnight incubation (37°C, 5% $CO_2$). The cell-seeded hydrogels were carefully transferred to a new 24-well plate to minimise risks of cell attachment to the well surface and to ensure that only the cells attached to the hydrogel were cultured further. 2 ml of cell culture medium were added to each sample-containing well, and the plate was incubated for up to 7 days (37°C, 5% $CO_2$), with the cell culture medium replaced every two days. After incubation, samples (n = 6) were washed with PBS (×3) and transferred to a new 24-well plate, to ensure that only cells attached to and grown on the hydrogel were imaged and respective metabolic activities quantified. Subsequently, the dying agents of Calcein AM and Ethidium homodimer-1 were added, and the sample plate incubated in the dark for 20 minutes. Live/dead stained hydrogels were placed onto a glass slide for fluorescence microscopy imaging (Leica DMI6000 B). Cells grown on tissue culture treated plastics were used as the positive control (Nunc, UK). Other than live/dead staining, cell viability was assessed at selected time points (day 1, 4 and 7) using Alamar Blue assay (ThermoFisher Scientific, UK) according to the manufacturer's guidance. Six replicates were used at each time point for each hydrogel group. VP-SEM (Hitachi S-3400N VP) combined with Deben cool stage control (Model: LT3299) was also employed to observe cell attachment after 7-day culture.

**2.11 Subcutaneous stability test *in vivo***

All animal studies were conducted under procedures approved by the University of Leeds Ethics Committee and under the UK Home Office project license (PPL: 70/8549). Hydrogels 4VBC-MA* (Ø: 15 mm, n=2) were prepared in a PBS solution supplemented with 1 wt.% I2959. The UV-cured hydrogels were washed in distilled water and dehydrated in an ascending series of distilled water-ethanol mixtures and air-dried. Air-dried samples of 4VBC-MA* (n=2) were disinfected in a cell culture hood via incubation in a 70 vol.% ethanol solution under UV exposure (as reported in section 2.10). Ethanol-disinfected samples were incubated in a sealed PBS-containing vial for 24 hours, prior to subcutaneous implantation in a 3-month old Sprague Dawley (SD) male rat (300-350 g). The rat was placed in a trifluorane chamber and anaesthetised (level 5 trifluorane and 2.5% oxygen) for ~2 minutes. After induction, the animal was transferred onto a heated mat and anaesthesia was maintained (level 2.5 triflurane and 2.5% oxygen) via a nose cone. The upper and lower sections of the back were shaved and two full thickness skin incisions were made per section. The subcutaneous tissues were then bluntly dissected using artery forceps to create two pockets so that the PBS-swollen hydrogels were individually implanted within these pockets. The wounds were closed using 5-0 ethilon sutures. The wound was cleaned with sterile injection water and the triflurane switched off and the oxygen left on until the rats were fully recovered. A buprenorphine (0.03 mg·ml$^{-1}$, 300 μl) injection was given to the rat. Following 4-week implantation, the rat was sacrificed using a carbon dioxide chamber for 6 minutes, followed by cervical dislocation. The samples and the surrounding tissue were then dissected out, photographed and fixed in 10% neutral buffered formalin (NBF).

**2.12 *In vivo* GBR pilot study**

A 30-day proof-of-concept study *in vivo* was carried out in a rat calvarial defect model using eight 3-month-old Sprague Dawley rats. The objectives of this study were to corroborate the results obtained *in vitro* and to build the foundation for a future larger animal study, de-risking animal testing in the early stage of prototype development, in line with the principles of the 3Rs (Replacement, Reduction and Refinement). The experimental GBR procedure was performed as previously reported [50,51]. Two full-thickness calvarial bone defects (Ø: 5 mm) were created with a trephine. A sample of the commercial membrane (Bio-Gide®, Geistlich) was trimmed to 6 mm diameter and always implanted to the intracranial side of the defect, as one of the leading GBR membrane gold standards. A commercial sample of deproteinised bovine bone mineral graft (Bio-Oss®, Geistlich) was carefully applied to fill the bone defect area, whereby a negative control group without Bio-Oss® was also used. Subsequently, either the ethanol-disinfected hydrogel 4VBC-MA* (Ø: 8 mm, UV-cured in I2959-supplemented PBS solution) or Bio-Gide® (as control, Ø: 8 mm) was applied to the extracranial side to assess the barrier functionality in the most degradative environment of the presented GBR model [14]. Finally, the flap was sutured in layers to close the wound. The commercial membranes were always oriented with the rough surface towards the bone defect. Two rats of each group were sacrificed by schedule 1 method at both 1 and 4 weeks, and the samples were removed for histological analysis. The experimental groups are presented in Table 1.

**Table 1.**

### 2.12.1 Histology

The rat calvarium samples (including the materials) were fixed in 10% NBF. Following decalcification, the specimens were embedded in paraffin for sectioning. Anterior–posterior consecutive sections of 8–10 μm every 100 μm were cut with a fully motorised and programmable rotary microtome (Leica RM 2265) and stained with haematoxylin and eosin (H&E) according to the experimental groups presented in Table 1.

### 2.13 Statistical analysis

Statistical analysis was carried out using OriginPro 8.51. Statistical differences were determined by one-way ANOVA and the *post hoc* Tukey test. A *p* value less than 0.05 was considered to be significantly different. Data are presented as mean ± SD.

## 3. Results and Discussion

The fabrication of a GBR membrane was pursued via sequential functionalisation of AC with 4VBC and MA residues, followed by solubilisation in an I2959-supplemented aqueous solution and UV curing (Figure 1). Coupling of both 4VBC and MA residues to lysine functions was hypothesised to generate a mechanically competent UV-cured material with enhanced enzymatic stability, reduced swelling ratio and barrier functionality, due to the nearly complete functionalisation and crosslinking of AC's primary amino groups, and the capability of 4VBC's aromatic residues to chelate and deactivate collagen-degrading MMPs *in vivo* [44]. The characterisation of the AC products will be presented at molecular, micro and macroscopic scales, and following L929 cell culture *in vitro*. Ultimately, a proof-of-concept study *in vivo* in both a subcutaneous implantation model and a GBR model in rats will also be discussed to explore the material degradability and tissue formation following material implantation.

**Figure 1.**

### 3.1 Chemical characterisation of sequentially functionalised AC

UV-cured collagen hydrogels have been shown to display adjusted macroscopic properties depending on the degree and type of covalent functionalisation of respective hydrogel-forming collagen precursors [30,46,52]. For example, an increase in the degree of methacrylation was demonstrated to directly correlate with the gel content (as an indirect measure of molecular crosslinking density), as well as the compression modulus and proteolytic stability (at the macroscopic scale) of respective UV-cured AC hydrogel [43]. Given the relevance of aforementioned macroscopic properties on GBR functionalities, functionalisation of AC was therefore pursued via sequential reaction with both 4VBC and MA ([monomer]·[Lys]$^{-1}$ =25), aiming to accomplish nearly complete derivatisation of AC's lysines and exploit the relatively high reactivity of MA [47], as well as the molecular rigidity and metal-chelation capability of 4VBC [44]. Both TNBS and ninhydrin assays have been reported to be reliable methods to quantify the molar content of free amino groups [20,34,47] and were therefore selected to indirectly determine the degree of functionalisation in obtained AC products. Both assays confirmed a decreased molar content of free primary amino groups in 4VBC-reacted AC ([Lys]= 2.89 ± 0.06 → 2.38 ± 0.02 mol·g$^{-1}$), corresponding to an averaged degree of functionalisation of 18 mol.% (Table 2), in line with previous reports [43,44].

**Table 2.**

The relatively large molar content of free primary amino groups measured in the collagen product therefore supported the use of 4VBC as the reacting monomer of the first functionalisation step (prior to the reaction with MA), in line with the decreased reactivity of 4VBC towards nucleophilic substitution [43,45,46]. When the 4VBC-functionalised AC precursor was reacted with MA ([MA]·[Lys]$^{-1}$=25), nearly 100 mol.% of AC functionalisation was observed by TNBS and ninhydrin assays (Table 2), indicating successful covalent coupling of both monomers and an increased molar fraction of MA compared to 4VBC residues in the retrieved AC product. The lower degree of AC functionalisation with the aromatic (4VBC) compared to the methacrylate (MA) monomer is attributed to the generation of hydrochloric acid during the reaction of AC's primary amino groups with 4VBC's methylene bridges, leading to protonation and decreased reactivity of remaining primary amino groups (Figure S1).

In contrast to previous reports [33,34,35,43], this sequential functionalisation strategy therefore proved successful to enable nearly-complete functionalisation of AC amino group terminations with two distinct monomers, promising a wider range of macroscopic and degradation profiles in the absence of dispersed [6,41] or crosslinked [37,39] synthetic phase.

Together with the confirmation of sequential functionalisation, the reaction product of AC was characterised by CD spectroscopy to elucidate the impact of sequential functionalisation on AC's triple helix organisation (Figure 2). Dichroic patterns of sample 4VBC-MA proved to be comparable to the ones of native AC, whereby a negative peak at about 198 nm was observed corresponding to the presence of left-handed polyproline II helices; as well as a positive peak at about 220 nm, attributed to right-handed triple helices.

**Figure 2.**

Aforementioned CD spectra proved to be in stark contrast with the CD spectrum of gelatin [45,46], which shows no positive CD peak and confirming the presence of a randomly oriented polypeptide configuration. Other than the CD patterns, the magnitude ratio between positive and negative peak intensities (*RPN*) was also calculated and found to be equal to 0.12 and 0.088 in the case of native AC and SAP, respectively. While the former value appears to be in line with previous reports on type I collagen [30,43,45,46], the decrease of *RPN* in the SAP suggests a reduction in the content of AC triple helices. This observation is likely due to the nearly complete derivatisation of triple helix-stabilising primary amino groups of lysine terminations [47,53] and to the partial introduction of bulky 4VBC aromatic groups (Table 2).

### 3.2 Fabrication of UV-cured hydrogels

When 4VBC-MA was dissolved in I2959-supplemented aqueous solutions, full gelation was accomplished following exposure to UV light, an observation that supported the synthesis of a covalently crosslinked network of sequentially functionalised AC triple helices. No detectable macroscopic differences were observed between hydrogels 4VBC-MA* and 4VBC* (Figure S2, Supp. Inf.), and with respect to state-of-the-art glutaraldehyde crosslinked collagen [54]. Cool-stage SEM of hydrogels 4VBC-MA* revealed a homogeneous distribution of interconnected micropores (Ø ≤10 µm) across the material (Figure 3).

**Figure 3.**

High pore interconnectivity has been observed in previously reported UV-cured networks of covalently crosslinked collagen molecules [30,33,35,46], and indirectly suggests densification of collagen molecules via covalent crosslinking as well as improved degradability and mechanical properties [36]. Furthermore, control of internal pore size can also be leveraged to minimise risks of soft tissue cell infiltration in GBR membranes, as supported by the low porosity revealed by Bio-Gide®, as one of the current clinical gold standards (Figure S3, Supp. Inf.). Typically, micro-pores (Ø: 5-20 µm) are beneficial to limit the passage of cells during the initial phase of bone healing, while larger pores formed following initiation of the membrane degradation process may support bone growth in the degrading membrane [10,55,56]. The aforementioned observation of relatively small internal pores therefore provides the above hydrogel with inherent soft tissue barrier functionality, given the confined space available to cells, which limits cell spreading, proliferation and vascularisation [33, 57]. The importance of pore size in barrier functionality is also in agreement with the low porosity depicted by SEM in the sample of commercial collagen membrane used for GBR (Figure S3, Supp. Inf.).

Given the secondary interaction capability of AC amino acidic terminations and grafted photoactive residues, the effect of the UV-curing solvent on hydrogel properties was investigated as a new strategy to achieve hydrogel controllability. To explore the gelation kinetics and confirm the formation of a UV-induced hydrogel network, the SAP was dissolved in either 10 mM HCl (pH ~2), 17.4 mM AcOH (pH ~3) or 10 mM PBS (pH 7.5), as typical solvents for AC. Respective solutions were tested with oscillatory time sweeps during exposure to UV light, where significant increase in storage and loss moduli was promptly observed in all aqueous systems once UV light was activated (Figure 4). The value of *G'* was significantly (> 175-fold) higher than the one of *G''* after 180 s of UV irradiation (Table S1, Supp. Inf.), indicating a predominantly elastic behaviour as expected by the formation of a UV-induced covalently crosslinked collagen network. Following complete gelation, hydrogels 4VBC-MA* revealed a 20-fold increase in storage modulus compared to hydrogel 4VBC* [43], indicating significantly increased hydrogel elasticity, which is key to ensure enhanced clinical handling of the GBR membrane, and to minimise risks of mechanical damage following membrane implantation *in vivo*. The aforementioned variation in shear modulus can be rationalised considering the significant difference in the degree of functionalisation between sequentially and 4VBC-functionalised AC precursors (*F*: 18 → 95 mol.%, Table 2). These results therefore confirm the direct effect of this molecular parameter on the macroscopic properties of respective UV-cured hydrogels, as previously observed with single functionalised collagen precursors [43,46,51].

**Figure 4.**

Although the gelation profiles were found to display comparable trends, it was interesting to measure quantitative differences in shear moduli depending on the UV-curing solvent used for hydrogel network formation. When dissolved in 10 mM HCl, sample 4VBC-MA revealed the highest values of storage and loss modulus during the entire time sweep, whereby the value of *G'* (8149 ± 345 Pa) was found to be more than 200 times higher than the value of *G"* (40 ± 12 Pa) following UV treatment (Figure 4A). The hydrogels 4VBC-MA* therefore revealed a 20-fold increase in storage modulus compared to hydrogel 4VBC*, i.e. obtained via UV-curing AC precursors bearing 4VBC residues only [43]. These observations therefore demonstrate the significantly increased elasticity of UV-cured sequentially functionalised AC hydrogels, which is key to enable hydrogel applicability in GBR therapy. The aforementioned values of storage modulus measured in hydrogels 4VBC-MA* proved to be higher than the ones reported with both orthogonally crosslinked collagen-hyaluronic acid co-networks [37] and carboxymethyl cellulose-crosslinked collagen [40], and close to the ones of nanostructured collagen-PLGA composites [41], whilst higher values were revealed when type I collagen hydrogels were crosslinked with high molecular weight dextran [36]. These results therefore indicate the significant effect of the presented sequential functionalisation route on the mechanical properties of collagen-based materials prepared in the absence of either secondary polymers or collagen self-assembling step.

Other than the rheological differences at equilibrium, the solution of sample 4VBC-MA in 10 mM HCl displayed a constantly higher value of *G'* with respect to the *G"* prior to, and no cross-over point following, UV irradiation. This observation suggests the development of secondary interactions and molecular entanglements between sequentially functionalised AC molecules, such as aromatic-aromatic and hydrogen-bonding interactions, which is in agreement with the relatively high solution acidity (pH= 2), the high concentration of AC in the photoactive solution (1.2 wt.%) and the presence of aromatic and methacrylate residues covalently coupled to AC. In contrast to the rheograms recorded in 10 mM HCl solution, samples of 4VBC-MA dissolved in either 17.4 mM AcOH or 10 mM PBS revealed a clear cross-over point following UV activation, which was reached more quickly in the former (~ 4.5 s) compared to the latter (~ 7.5 s) solvent. UV-curing in diluted acetic acid also proved to generate covalent networks with significantly increased storage and loss moduli (*G'*= 3675 ±185 Pa; *G"*= 19 ± 1 Pa) with respect to UV-curing of the PBS-dissolved SAP (*G'*= 796 ± 61

Pa; *G''* = 5 ± 1 Pa) (Figure 4B and C).

As a well-known ampholyte, environmental factors, such as pH, ionic strength and salt concentration, as well as the amino acidic composition, affect secondary interactions and long-range configuration of collagen molecules [34,40, 58]. For instance, incubation of crosslinked type I collagen in aqueous solutions of decreased acidity was exploited to accomplish hydrogels with decreased swelling by controlling the electrostatic interactions between amino acidic side chains [59]. In this study, the sequential functionalisation with both 4VBC and MA residues is expected to contribute new secondary interactions between AC molecules. This observation is supported given the pH-dependent capability of aforementioned aromatic and methacrylate groups to mediate $\pi$-$\pi$ stacking interactions and hydrogen bonds, respectively, the nearly complete consumption of AC's free amino groups and the consequent decrease in AC's isoelectric point. Indeed, the isoelectric point of collagen was found to decrease from ~6.5 to ~5.5 following functionalisation of 30 mol.% of lysines with methacrylate residues [34], in agreement with the increased molar content of negatively charged carboxylic functions in the collagen backbone. Therefore, the increased molar content of MA (compared to 4VBC) residues and the decreased isoelectric point in the SAP are attributed to explain the above-mentioned decrease of storage and loss moduli in AC networks prepared in solutions of increased pH. With the increase in pH, the presence of OH$^-$ ions generates weaker hydrogen bonds [60] as well as electrostatic repulsion between AC side chains terminating with deprotonated carboxylic terminations [34], resulting in decreased shear moduli, as observed with UV-cured MA-functionalised AC [43]. Unlike the case where UV curing was carried out at increased pH (Figure 4 B-C), significant secondary interactions could be detected when the sample 4VBC-MA was solubilised in the 10 mM HCl solution (pH ~ 2), whereby no cross-over point and a predominantly elastic response was measured following activation of UV light (Figure 4A).

### 3.3 Physical characterisation: swelling ratio, gel content and compression properties

AC networks UV-cured in either 10 mM HCl, 17.4 mM AcOH or 10 mM PBS were further characterised *in vitro* with respect to their crosslinked architecture, macroscopic properties and enzymatic degradability, aiming to assess their applicability as GBR membrane. A high gel content ($G \geq 95$ wt.%) was exhibited by all dry AC networks following 24-hour incubation in 17.4 mM AcOH (Figure 5), in agreement with the nearly complete functionalisation of AC and in contrast to the complete dissolution observed with a control sample of native AC in the same conditions. Although UV curing in the PBS solution proved to generate networks with high gel content ($G= 95\pm3$ wt.%), these values proved to be lower than the ones measured in the two acid-cured groups, an observation that agrees with previously observed solvent-induced variations in gelation kinetics (Figure 4) and shear moduli (Table S1, Supp. Inf.). Since the gel content is directly related to the extent of conversion of the photoactive precursor to the crosslinked network [61], the above-mentioned values of $G$ therefore provide additional confirmation of the successful UV-induced crosslinking reaction between sequentially functionalised AC molecules in all three solvents. This is key aiming to minimise risks of releasing soluble collagen fraction following contact with the biological environment, as reported with currently available collagen membranes [62], and to accomplish clinically-relevant elasticity and swelling properties. The elasticity of the GBR membrane is indeed of utmost importance from a clinical point of view to enable clinicians to accomplish membrane manipulation and adaptation to the bone defects and to allow for proper fixation of the barrier membrane when used alone or in combination with a bone graft.

Given that water acts as a plasticiser in biopolymers [32], it is critical that respective water-induced swelling does not deteriorate the clinical handleability of the collagen membrane and that the membrane dimensional conformation with the tissue defect is ensured. Indeed, a relatively low swelling ratio in the GBR membrane is important for clinicians to ensure primary wound closure. Conversely, an unexpected /uncontrolled swelling ratio during the early healing phase might lead to membrane /graft exposure and healing by secondary intention, which would compromise the regeneration outcome. Consequently, it was important to measure the swelling behaviour of accomplished UV-cured hydrogel networks. All dry collagen samples took up, and swelled when incubated in, distilled water, whereby a swelling ratio (*SR*) in the range of 350±48–684±76 wt.% was measured upon solvent equilibration (Figure 5). Especially the *SR* of the PBS-cured hydrogels was found to be comparable to the one of Bio-Gide® (*SR*= 289±22 wt.%, p= 0.0602) and significantly smaller than the one measured in crosslinked collagen systems intended for GBR therapy [38,42,63] as well as UV-cured homo-networks of either 4VBC- or MA-functionalised collagen [43,44,46]. From a physical property point of view, the aforementioned swelling behaviour therefore support the GBR applicability of this hydrogel, aiming to reduce the volumetric expansion of the membrane at the defect site *in vivo* following contact with the physiological medium and to ensure tissue occlusivity [14,49]. The aforementioned relatively low values of *SR* can be attributed to the significant crosslink density in the resulting UV-cured network and the relatively high concentration of collagen used in the UV-curing aqueous system.

**Figure 5.**

Although it is difficult to confirm the exact nature of the covalent crosslinks formed in the UV-cured network, i.e. homo-crosslinks (either 4VBC-4VBC or MA-MA), hetero-crosslinks (4VBC-MA) or combinations thereof, the decreased values in *SR* provides indirect evidence of complete crosslinking between the photoactive residues covalently coupled to the same AC backbone. These results therefore support the direct effect of the degree functionalisation on the macroscopic properties of UV-cured sequentially functionalised AC hydrogels.

Samples prepared in the 17.4 mM AcOH solution (pH ~ 3) revealed the largest swelling ratio (*SR*), whilst insignificant differences of *SR* were recorded in samples cured in solutions of either 10 mM HCl or 10 mM PBS. Typically, networks with increased gel content tend to display decreased swelling properties, in light of the increased content of crosslinks between polymer chains. Whilst the above data seem to deviate from the inverse *G-SR* relationship, electrostatic effects are likely to play an additional contribution in this case [64], given the amphoteric character of AC and the nearly full consumption of primary amino groups in the SAP. The swelling of collagen is known to decrease in neutral pH, due to the protonation of, and electrostatic interaction between, amino and carboxylic terminations, whilst increased swelling is observed in acidic or basic environment due to the electrostatic repulsions between the amino acidic side groups of collagen [40,59,65,66,67]. Consequent to the full consumption of primary amino groups, the SAP dissolved in the PBS-based UV-curing system is only expected to expose fully deprotonated carboxylic terminations. The most likely explanation for the significantly lower values of *SR* in hydrogels UV-cured in physiological (10 mM PBS, pH 7.5) with respect to acidic (e.g. 17.4 mM AcOH, pH 3) conditions is therefore attributed to the secondary interactions between ionic species of PBS and the deprotonated carboxylic terminations of the SAP. The aforementioned secondary interactions are unlikely to be found when the AC network is prepared in either 10 mM HCl or 17 mM AcOH, since ionisation of AC's carboxylic terminations ($pK_a$= 3.65-4.25) is decreased in these aqueous media (pH ≤ 3). Consequently, this solvent-induced mechanism of swelling ratio control may offer an additional dimension to build hydrogels with controlled structure-function relations.

Given the large swelling ratio in physiological conditions and the fact that water acts as a plasticiser for collagen, the attention moved to the characterisation of the mechanical properties in the water-equilibrated collagen networks. Samples with a diameter/height ratio of ~2 were prepared to minimise risks of sample tilting during compression, which is typically observed when hydrogel samples with a relatively large height are measured. All samples described a *J*-shaped stress-compression curve (Figure 6A), which is typical of biological tissues [36,43,49,66] and which is attributed to the complex interplay of aqueous fluid and covalently-crosslinked collagen molecules under compression [68].

**Figure 6.**

The compression properties of the AC networks varied according to the UV-curing solvent, whereby the PBS-cured samples displayed the largest compressibility (up to nearly 90% compression) and significantly reduced compression modulus ($E_c$= 22±9 kPa), compared to both acid-cured samples ($E_c$ = 155±24–178±20 kPa) (Figure 6B). This variation in compression modulus agrees with previously discussed trends in shear moduli (Figure 4), confirming the suitability of selected sample geometry and compression setup, so that a multimodal mechanical characterisation was accomplished. The variation in compression modulus also reflects the increased compressibility and decreased gel content measured in AC networks UV-cured in PBS (Figure 5). The increased compressibility presented by the PBS-cured hydrogel networks is key aiming to accomplish safe membrane implantation and fixation at the critical-sized defect site, minimising risks of membrane collapse in the defect and /or membrane damage, including curling or wrinkling observed following contact with body fluid [16].

### 3.4 Proteolytic stability *in vitro* and subcutaneous implantation *in vivo*

Samples UV-cured in PBS were subsequently selected for degradation tests in collagenase media, given their relatively low swelling ratio, relatively high compressibility and acid-free synthesis. These three features were considered key for GBR therapy, aiming to ensure occlusivity, elasticity and surgical handling, as well as cell tolerability and safety *in vivo*, respectively. Following 14-day enzymatic incubation (37 ºC, 1 CDU·ml$^{-1}$), samples 4VBC-MA* successfully displayed significantly higher relative mass at both days 4 and 14 ($\mu_{rel}$ = 73±8 wt.% → 50±4 wt.%), with respect to both controls of UV-cured 4VBC-functionalised homo-network ($\mu_{rel}$ = 46±2 → 24±4 wt.%) and the commercial collagen membrane Bio-Gide® ($\mu_{rel}$ = 19±5 → 14±4 wt%) (Figure 7A). The enhanced proteolytic stability of the UV-cured samples 4VBC-MA* was further confirmed in solutions supplemented with increased collagenase concentration (2.5 CDU·ml$^{-1}$), whereby nearly 50 wt.% of the original mass was still retained ($\mu_{rel}$ = 57±2 → 47±3 wt.%), whilst almost all sample of Bio-Gide® was lost ($\mu_{rel}$ = 2±4 → 1±2 wt.%) (Figure 7B). These gravimetric results were in agreement with the digital macrographs captured during sample incubation *in vitro*, whereby the hydrogel 4VBC-MA* revealed minimal changes in macroscopic integrity in contrast to the samples of the collagen membrane (Figure 7C). The aforementioned degradation trends were also confirmed when control samples of native AC were used, which revealed a 10-fold decrease in relative mass with respect to the UV-cured samples 4VBC-MA* following 4-day incubation with increased collagenase concentration (2.5 → 5 CDU ml$^{-1}$, Figure S4, Supp. Inf.).

**Figure 7.**

Similar trends were also exhibited by the homo-networks of both 4VBC* ($p$< 0.005), whereby nearly 2-fold decrease in relative mass was recorded with respect to samples 4VBC-MA* (Figure S4, Supp. Inf.). These results provide additional confirmation that the coupling of both aromatic *and* methacrylate residues onto AC's lysines and consequent nearly complete functionalisation of AC lysines are key to accomplish long-lasting proteolytic stability of the AC hydrogels. The long-lasting proteolytic stability of hydrogel 4VBC-MA* observed *in vitro* was ultimately corroborated by an *in vivo* subcutaneous implantation test in rats, whereby samples successfully demonstrated to remain intact and to display only low diameter reduction (~ 13%) after 4 weeks *in vivo* (Figure 7D). The results of this test *in vivo* therefore contributed additional research leverage aiming to progress preliminary prototype evaluation with a proof-of-concept study in a GBR defect model *in vivo*, minimising animal testing risks and respective ethical issues (section 3.6).

Overall, the presence of 4VBC-crosslinked AC molecules is expected to induce cation-$\pi$ interactions between 4VBC's aromatic rings and active zinc sites of collagenases, so that drug-free enzymatic deactivation can be successfully accomplished with minimal peptide bond scission [43]. Together with 4VBC, the presence of MA in the SAP is on the other hand beneficial to accomplish increased crosslink density following UV curing and delayed AC degradation. The 14-day proteolytic stability ($\mu_{rel}$ = 47±3 wt.%) of 4VBC-MA* *in vitro* was significantly higher than that reported for hyaluronic acid-collagen co-networks [37], UV-cured methacrylated collagen [34], as well as PVA- [66], genipin- [67] and PEG-crosslinked collagen [69], whilst, as expected, it was lower than that of carboxymethyl cellulose-crosslinked collagen [42], in light of the protease insensitivity of cellulose.

**3.5 Biocompatibility and cell proliferation *in vitro***

Consequent to the degradation tests, efforts focused on assessing the tolerability of the accomplished AC hydrogel 4VBC-MA* with L929 fibroblasts, aiming to confirm cell viability, the absence of any hydrogel-induced toxic effects, and cell barrier functionality (Figure 8). Accordingly, dry UV-cured samples were incubated in a 70 vol.% ethanol solution under UV exposure, to achieve material disinfection [70,71]. Live/dead staining of the L929 fibroblasts cultured on to the resulting UV-cured samples 4VBC-MA* (Figure 8A) revealed high cellular tolerability of the hydrogel (Figure 8B), whereby only partially detectable dead cells were observed (in red) at the end of day 7 (Figure 8C). These observations are in agreement with the previously reported biocompatibility of AC homo-networks 4VBC* and MA* [43], and the high gel content values measured on the hydrogel 4VBC-MA* ($G$= 95±3 wt.%, Figure 5), which indicate full network formation and minimal presence of unreacted species following UV-curing. The post-synthesis hydrogel dehydration in aqueous solutions of increasing ethanol concentration is also expected to serve as an additional hydrogel purification step, whereby any unreacted residues are washed away from the samples, reducing risks of toxic response during cell culture. Superimposed confocal microscopy images of freshly-synthesised and 7-day cultured hydrogels also revealed that the viable cells were mainly localised on the surface of, rather than across, the hydrogel structure. This surface-localised growth of L929 fibroblasts is in agreement with the presence of relatively small micro-pores observed in the hydrogel microstructure (Figure 3). This, together with the fact that no detection of dead (red) cells was observed (Figure 8D), therefore supports both the barrier functionality and high tolerability of this hydrogel.

**Figure 8.**

Cell viability results were corroborated via the Alamar Blue assay, which was carried out during the 7-day culture to gain insight on the cell metabolic activity (Figure 9A). The cellular activity proved to be directly related to the culture time on both the collagen hydrogel and the tissue culture plastic (TCP) control. The cells cultured on TCP exhibited a 3- and 16-fold averaged increase in the metabolic activity at day 4 and 7 of culture, respectively, when compared to the metabolic activity of cells at day 1. Interestingly, the cells cultured on the AC hydrogel exhibited a significantly increased metabolic activity when compared to the TCP control. Cells cultured on the hydrogels exhibited a 15- and 40-fold averaged increase in metabolic activity at day 4 and 7, respectively, when compared to day 1. These findings therefore further evidence the viability of hydrogel-cultured cells (Figure 9B) and indicate an increased rate of cell proliferation induced by the AC hydrogel when compared to conventional TCP.

**Figure 9.**

Consequently, the surface of the AC hydrogel effectively supports the migration and proliferation of L929 cells, in line with the effect of collagen on cell adhesion, proliferation and survival *in vitro* [72], and despite the sequential functionalisation and UV-curing route pursued for the hydrogel fabrication. At the same time, the presence of the relatively small micropores in the hydrogel (Figure 3) is seen to prevent cell spreading and proliferation in the inner hydrogel structure (Figure 8D). This explanation is supported by the significantly less porous hydrogel surface depicted at day 7 of cell culture (Figure 9C-D) compared to the native hydrogel (Figure 3), proving cell attachment on the hydrogel surface. These observations therefore indirectly suggest the capability of the hydrogel to integrate with the surrounding tissue, yet minimising risks of tissue growth in the inner microstructure, in agreement with the functionality of a GBR membrane.

The proteolytic scission of peptide bonds in the biological environment may be associated with the formation of internal larger pores in the AC hydrogel at later time points (i.e. after 7 days of cell culture), raising risks of decreased barrier functionality. Conversely, a low decrease in sample diameter (~ 13%) was measured in the hydrogels following 4-weeks subcutaneous implantation *in vivo*, in line with the increased material integrity observed *in* *vitro* with respect to Bio-Gide® (Figure 7). These observations speak in favour of a surface erosion rather than bulk degradation mechanism, in agreement with previous degradation studies on collagen-based hydrogels [52,73]. According to a surface erosion mechanism, larger pores are predominantly formed in the outer layer rather than on the hydrogel core, generating minimal risks on the long-term barrier and tissue integration membrane functionalities.

The aforementioned data therefore demonstrate that the presented sequential collagen functionalisation route enables the simple fabrication of a biocompatible hydrogel with convenient microstructure for localised fibroblast growth and barrier functionality, yet avoiding complex manufacturing steps and multi-layered membrane architectures [15,16,19,42].

### 3.6 *In vivo* pilot study of GBR functionality

Having confirmed its biocompatibility and barrier functionality *in vitro*, as well as its 4-week material integrity *in vivo*, the attention focused on exploring the performance of the hydrogel as a GBR barrier membrane in a GBR animal model. Given the unknown tolerance *in vivo* of this previously-unreported AC membrane, a proof-of-concept study was carried out in rats using a relatively low number of animals, in line with the principles of the 3Rs for animal testing and aiming to lay down the research foundation for a future larger investigation *in vivo*. Samples of the AC hydrogel and Bio-Gide® were employed as the experimental group and the clinical gold standard, respectively. Calvarial defect models are often employed to assess the *in vivo* bone regeneration capacity, whereby a critical size defect is created in the exposed calvarial bone, treated with the experimental sample, followed by repositioning and suturing of the soft tissue [6,17,23,50]. Although this model offers a relatively easy surgical procedure to assess the material-induced bone regeneration, risks persist with respect to soft tissue cell infiltration from the underlying dura mater, especially when the membrane is not completely conformed to the surgical site. To minimise risks of soft tissue infiltration, a barrier GBR defect model was created in the calvaria of rats [74]. To assess the barrier functionality in the most degradative environment of the presented GBR defect [14], samples of 4VBC-MA* were always applied to the extracranial side (Figure 10A), i.e. facing the periosteum and soft tissue, whereby control groups were treated with Bio-Gide®. Compatibly with the nature of the proof-of-concept study, the intracranial side, i.e. facing the dura mater (Figure 10A), was on the other hand always covered with samples of Bio-Gide®. As one of the leading GBR membrane gold standards, Bio-Gide® has indeed been extensively documented in the literature to prevent the ingrowth of cells coming from the dura mater [75,76,77]. Ultimately, the effect of the graft material Bio-Oss® in the critical size defect was also explored.

**Figure 10.**

After one-week implantation *in vivo*, samples of Bio-Gide® (Figure 10B-E) and hydrogel 4VBC-MA* (Figure 10F-I) proved to act as a barrier regardless of the application of Bio-Oss®. The space between the two defect sides was still observed, whereby the formation of densely packed tissue was detected within the bone defect area and around the Bio-Oss® particles (Figure 10D-E & H-I, see black arrows). The distance between the two membranes proved to be decreased in the blank defects (Figure 10B-G) compared to defects filled with Bio-Oss® (Figure 10D-I), suggesting that the application of the membranes alone is insufficient to withstand the mechanical loads and preserve the integrity of the defect gap. Increased cell density was also observed in the graft-filled defects (Figure 10I, see black arrows), likely due to the fact that Bio-Oss® promotes osteoblast migration from the edges, towards the centre, of the defect, in line with the osteoconductive functionality of the selected graft material. Nevertheless, implantation of the membranes for one week *in vivo* may be too short for visualising any *de novo* bone tissue formation [78], as also supported by the presence of small gaps at the membrane-defect interface (Figure 10B-D & F-I). This observation is in line with the initial inflammation phase of the bone regeneration process [16], prior to the subsequent occurrence of angiogenesis, granulation tissue, and remodelling [79].

After four weeks of implantation, both samples of Bio-Gide® (Figure 11A-D) and hydrogel 4VBC-MA* (Figure 11E-H) remained intact, supporting the long-lasting proteolytic stability of the AC hydrogel previously observed *in vitro* and *in vivo* (Figure 7).

**Figure 11.**

The integrity of the sample of Bio-Gide® following four weeks *in vivo* is partially in contrast with previous results obtained *in vitro*, which indicated nearly complete dissolution of the commercial membrane following two-week incubation in a collagenase-supplemented medium (Figure 7A-C). This observation can be rationalised by considering the different enzymatic concentrations between *in vitro* and *in vivo* environments, as well as the relatively short duration (4 weeks) of this proof-of-concept study *in vivo*. Considering the typical bone regeneration process [13,80], an increased implantation time (12 weeks) may be needed in a follow on study to clearly demonstrate a compelling advantage in the degradation profile and GBR functionality of the AC prototype with respect to Bio-Gide®. The additional confirmation of proteolytic stability for at least four weeks in a GBR model *in vivo* is therefore key to de-risk future, larger investigations *in vivo* with the AC prototype prior to moving towards a first-in-human clinical evaluation.

The material integrity and the formation of an island of densely packed tissue in the centre of the four-week graft-free defect treated with Bio-Gide® and the hydrogel 4VBC-MA* (Figure 11E-F, see blue arrows) suggests the maintenance of the barrier function by both materials after four weeks as well as their integration with the surrounding bone. An increased cell density was apparent at the centre of the graft-free defects (Figure 11E-F), in line with the decrease in inter-membrane distance observed in the sample groups in no receipt of Bio-Oss®. The extensive presence of densely packed tissue was also detected after four weeks in the defect groups filled with Bio-Oss® and treated with either Bio-Gide® (Figure 11C-D) or hydrogel 4VBC-MA* (Figure 11G-H). Consequently, either the application of the osteoconductive graft material or the decrease in gap distance in the graft-free defects contributes to providing cells with the anchorage sites critical for cell attachment, proliferation and new ECM deposition. This result agrees with the safety, occlusivity, and soft tissue barrier functionality of the hydrogel measured *in vitro* (Figure 8 and Figure 9) and suggests that the sequential functionalisation and UV curing do not negatively impact on the biocompatibility of AC *in vivo*.

Overall, these preliminary *in vivo* results support the integrity and cell /tissue tolerability of the hydrogel following four-week implantation *in vivo*, as observed in the case of Bio-Gide®, and in line with enzymatic degradation (Figure 7) and barrier functionality (Figure 8) results obtained *in vitro*. Further larger scale follow on study should be carried out *in vivo* to assess hydrogel impact on bone regeneration and longer-term hydrogel stability in the GBR defect model, which can take advantage of more advanced 3D imaging technologies, such as micro-CT.

**Conclusions**

The sequential functionalisation of AC with two distinct ethylenically unsaturated monomers, i.e. 4VBC and MA, successfully enabled the fabrication of a UV-cured hydrogel network with competitive proteolytic stability, compressibility and swelling ratios with respect to the corresponding homo-networks, as well as increased proteolytic stability with respect to Bio-Gide®, as one of the current GBR collagen membrane gold standards. The selection of acidic or basic UV-curing aqueous systems proved to affect gelation kinetics, swelling ratio and compression properties, due to the effect of ionisable terminations of AC, covalently coupled monomers and the complete consumption of primary amino groups. As observed in collagenase-supplemented media *in vitro*, the UV-cured sequentially functionalised samples proved to remain intact for at least four weeks following subcutaneous implantation in rats, and to ensure space maintenance in a proof-of-concept study in a GBR model in rats. These *in vivo* results therefore provide compelling evidence aiming to de-risk future prototype development and support the additional large scale preclinical testing of this long-lasting AC membrane, following the principles of the 3Rs for experimental animal studies. A follow-on larger scale investigation *in vivo* will be key to confirm the longer-term stability of this material and its impact on bone regeneration, prior to moving forward with a first-in-human clinical evaluation.

**Declaration of Competing Interest**

G. Tronci, S.J. Russell and D.J. Wood are named inventors on a patent related to the fabrication of collagen-based materials.

**Acknowledgments**

The authors wish to thank the MRC Confidence in Concept scheme, the University of Leeds Medical Technologies IKC, the Clothworkers' Company and the EPSRC Impact Acceleration Account for financial support. Jackie Hudson and Dr. Sarah Myers are gratefully acknowledged for technical assistance with scanning electron microscopy and cell culture. The authors also wish to thank Dr. Roisin Holmes for assisting with the subcutaneous implantation study.

# References

[1] N. Mardas, J. Busetti, J.A. Poli de Figueiredo, L.A. Mezzomo, R. Kochenborger Scarparo, N. Donos. Guided bone regeneration in osteoporotic conditions following treatment with zoledronic acid. Clin. Oral Implants Res. 28 (2017) 362-371.

[2] Q. Yuan, L. Li, Y. Peng, A. Zhuang, W. Wei, D. Zhang, Y. Pang, X. Bi. Biomimetic nanofibrous hybrid hydrogel membranes with sustained growth factor release for guided bone regeneration. Biomater. Sci. 9 (2021) 1256-1271.

[3] I.A. Urban, A. Monje. Guided Bone Regeneration in Alveolar Bone Reconstruction. Oral Maxillofac. Surg. Clin. North. Am. 31 (2019) 331-338.

[4] J.J. El-Jawhari, K. Moisley, E. Jones, P.V. Giannoudis. A crosslinked collagen membrane versus a noncrosslinked collagen membrane for supporting osteogenic functions of human bone marrow multipotent stromal cells. Eur. Cells Mater. 37 (2019) 292-309.

[5] H. Owston, K.M. Moisley, G. Tronci, S.J. Russell, P.V. Giannoudis, E. Jones. Induced periosteum-mimicking membrane with cell barrier and multipotential stromal cell (MSC) homing functionalities. Int. J. Mol. Sci. 21 (2020) 5233.

[6] B. Li, Y. Chen, J. He, J. Zhang, S. Wang, W. Xiao, Z. Liu, X. Liao. Biomimetic Membranes of Methacrylated Gelatin/Nanohydroxyapatite/Poly(L-Lactic Acid) for Enhanced Bone Regeneration. ACS Biomater. Sci Eng. 6 (2020) 6737-6747.

[7] M.C. Bottino, V. Thomas, G. Schmidt, Y.K. Vohra, T.-M.G. Chu, M.J. Kowolik, G.M. Janowski. Recent advances in the development of GTR/GBR membranes for periodontal regeneration--a materials perspective. Dent. Mater. 28 (2012) 703-721.

[8] N. Donos, X. Dereka, N. Mardas. Experimental models for guided bone regeneration in healthy and medically compromised conditions. Periodontol. 2000 68 (2015) 99-121.

[9] M. Retzepi, N. Donos. Guided Bone Regeneration: biological principle and therapeutic applications. Clin. Oral Implants Res. 21 (2010) 567-576.

[10] M. Sanz, C. Dahlin, D. Apatzidou, Z. Artzi, D. Bozic, E. Calciolari, H. De Bruyn, H. Dommisch, N. Donos, P. Eickholz, J.E. Ellingsen, H.J. Haugen, D. Herrera, F. Lambert, P. Layrolle, E. Montero, K. Mustafa, O. Omar, H. Schliephake. Biomaterials and regenerative technologies used in bone regeneration in the craniomaxillofacial region: Consensus report of group 2 of the 15th European Workshop on Periodontology on Bone Regeneration. J Clin Periodontol. 46 (2019) 82-91

[11] M.J. Sánchez-Fernández, M. Peerlings, R.P. Félix Lanao, J.C.M.E. Bender, J.C.M. van Hest, S.C.G. Leeuwenburgh. Bone-adhesive barrier membranes based on alendronate-functionalized poly(2-oxazoline)s. J. Mater. Chem. B 9 (2021) 5848-5860.

[12] E. Costa e Silva, S.V. Omonte, A.G. Veiga Martins, H.H. Onibene de Castro, H.E. Gomes, É. Gonçalves Zenóbio, P.A. Dutra de Oliveira, M.C. Rebello Horta, P.E. Alencar Souza. Hyaluronic acid on collagen membranes: An experimental study in rats. Arch. Oral Biol. 73 (2017) 214-222.

[13] H. Guo, D. Xia, Y. Zheng, Y. Zhu, Y. Liu, Y. Zhou. A pure zinc membrane with degradability and osteogenesis promotion for guided bone regeneration: In vitro and in vivo studies. Acta Biomaterialia 106 (2020) 396-409.

[14] E. Calciolari, F. Ravanetti, A. Strange, N. Mardas, L. Bozec, A. Cacchioli, N. Kostomitsopoulos, N. Donos. Degradation pattern of a porcine collagen membrane in an in vivo model of guided bone regeneration. J. Periodontal Res. 53 (2018) 430-439.

[15] Y. Xue, Z. Zhu, X. Zhang, J. Chen, X. Yang, X. Gao, S. Zhang, F. Luo, J. Wang, W. Zhao, C. Huang, X. Pei, Q. Wan. Accelerated bone regeneration by MOF modified multifunctional membranes through enhancement of osteogenic and angiogenic performance. Adv. Health. Mater. 10 (2021) 2001369.

[16] Q. Wang, Y. Feng, M. He, W. Zhao, L. Qiu, C. Zhao. A hierarchical Janus membrane combining direct osteogenesis and osteoimmunomodulatory functions for advanced bone regeneration. Adv. Funct. Mater. 31 (2021) 2008906.

[17] H. Su, T. Fujiwara, K.M. Anderson, A. Karydis, M. Najib Ghadri, J.D. Bumgardner. A comparison of two types of electrospun chitosan membranes and a collagen membrane *in vivo*. Dent. Mater. 37 (2021) 60-70.

[18] J. Fang, R. Liu, S. Chen, Q. Liu, H. Cai, Y. Lin, Z. Chen, Z. Chen. Tuning the immune reaction to manipulate the cell-mediated degradation of a collagen barrier membrane. Acta Biomater. 109 (2020) 95-108.

[19] E. Prajatelistia, N.D. Sanandiya, A. Nurrochman, F. Marseli, S. Choy, D.S. Hwang. Biomimetic Janus chitin nanofiber membrane for potential guided bone regeneration applications. Carbohyd. Polym. 251 (2021) 117032.

[20] S. Federico, G. Pittaresi, F.S. Palumbo, C. Fiorica, F. Yang, G. Giammona. Hyaluronan alkyl derivatives-based electrospun membranes for potential guided bone regeneration: fabrication,

characterization and in vitro osteoinductive properties. Colloids Surf. B Biointerfaces 197 (2021) 111438.

[21] C.-C. Lin, J.-Y. Chiu. A novel γ-PGA composite gellan membrane containing glycerol for guided bone regeneration. Mater. Sci. Eng. C 118 (2021) 111404.

[22] M. Wu, Z. Han, W. Liu, J. Yao, B. Zhao, Z. Shao, X. Chen. Silk-based hybrid microfibrous mats as guided bone regeneration membranes. J. Mater. Chem. B 9 (2021) 2025-2032.

[23] L. Ebrahimi, A. Farzin, Y. Ghasemi, A. Alizadeh, A. Goodarzi, A. Basiri, M. Zahiri, A. Monabati, J. Ai. Metformin-Loaded PCL/PVA Fibrous Scaffold Preseeded with Human Endometrial Stem Cells for Effective Guided Bone Regeneration Membranes. ACS Biomater. Sci. Eng. 7 (2021) 222-231.

[24] M.B. Bazbouz, H. Liang, G. Tronci. A UV-cured nanofibrous membrane of vinylbenzylated gelatin-poly(ε-caprolactone) dimethacrylate co-network by scalable free surface electrospinning. Mater. Sci. Eng. C 91 (2018) 541-555.

[25] S.K. Boda, N.G. Fischer, Z. Ye, C. Aparicio. Dual oral tissue adhesive nanofiber membranes for pH-responsive delivery of antimicrobial peptides. Biomacromolecules. 21 (2020) 4945-4961.

[26] X. Niu, L. Wang, M. Xu, M. Qin, L. Zhao, Y. Wei, Y. Hu, X. Lian, Z. Liang, S. Chen, W. Chen, D. Huang. Electrospun polyamide-6/chitosan nanofibers reinforced nano-hydroxyapatite/polyamide-6 composite bilayered membranes for guided bone regeneration. Carbohyd. Polym. 260 (2021) 117769.

[27] E. Choi, S. Bae, D. Kim, G.H. Yang, K. Lee, H.-J. You, H.J. Kang, S.-J. Gwak. S.H. An, H. Jeon. Characterization and intracellular mechanism of electrospun poly(ε-caprolactone) (PCL) fibers incorporated with bone-dECM powder as a potential membrane for guided bone regeneration. J. Ind. Eng. Chem. 94 (2021) 282-291.

[28] G. de Souza Balbinot, E. Antunes da Cunha Bahlis, F. Visioli, V.C. Branco Leitune, R.M. Duarte Soares, F. Mezzomo Collares. Polybutylene-adipate-terephtalate and nobium-containing bioactive glasses composites: development of barrier membranes with adjusted properties for guided bone regeneration. Mater. Sci. Eng. C 125 (2021) 112115.

[29] G. Tronci. The application of collagen in advanced wound dressings, in: S. Rajendran (Ed.), Advanced Textiles for Wound Care, second ed., Elsevier Ltd., 2019, pp. 363-389.

[30] R. Holmes, S. Kirk, G. Tronci, X. Yang, D. Wood. Influence of telopeptides on the structural and physical properties of polymeric and monomeric acid-soluble type I collagen. Mater. Sci. Eng. C 77 (2017) 823-827.

[31] M.G. Haugh, C.M. Murphy, R.C. McKiernan, C. Altenbuchner, F.J. O'Brien. Crosslinking and mechanical properties significantly influence cell attachment, proliferation, and migration within collagen glycosaminoglycan scaffolds. Tissue Eng. Part A 17 (2011) 1201-1208.

[32] G. Tronci, A.T. Neffe, B.F. Pierce, A. Lendlein. An entropy–elastic gelatin-based hydrogel system. J. Mater. Chem. 20 (2010) 8875-8884.

[33] J. Yang, Y. Li, Y. Liu, D. Li, L. Zhang, Q. Wang, Y. Xiao, X. Zhang. Influence of hydrogel network microstructures on mesenchymal stem cell chondrogenesis *in vitro* and *in vivo*. Acta Biomaterialia 91 (2019) 159-172.

[34] J. Yang, Y. Xiao, Z. Tang, Z. Luo, D. Li, Q. Wang, X. Zhang. The negatively charged microenvironment of collagen hydrogels regulates the chondrogenic differentiation of bone marrow mesenchymal stem cells *in vitro* and *in vivo*. J. Mater. Chem. B 8 (2020) 4680-4693.

[35] K.E. Drzewiecki, A.S. Parmar, I.D. Gaudet, J.R. Branch, D.H. Pike, V. Nanda, D.I. Shreiber. Methacrylation induces rapid, temperature-dependent, reversible self-assembly of type-I collagen. Langmuir 30 (2014) 11204-11211.

[36] X. Zhang, Y. Yang, J. Yao, Z. Shao, X. Chen. Strong collagen hydrogels by oxidised dextran modification. ACS Sustain. Chem. Eng. 2 (2014) 1318-1324.

[37] F. Chen, P. Le, G. M. Fernandes-Cunha, S. C. Heilshorn, D. Myung. Bio-orthogonally crosslinked hyaluronate-collagen hydrogel for suture-free corneal defect repair. Biomaterials 255 (2020) 120176.

[38] Y. Wei, Y.-H. Chang, C.-J. Liu, R.-J. Chung. Integrated Oxidized-Hyaluronic Acid/Collagen Hydrogel with β-TCP Using Proanthocyanidins as a Crosslinker for Drug Delivery. Pharmaceutics 10 (2018) 37.

[39] R. Holmes, X.B. Yang, A. Dunne, L. Florea, D.J. Wood, G. Tronci. Thiol-Ene Photo-Click Collagen-PEG Hydrogels: Impact of Water-Soluble Photoinitiators on Cell Viability, Gelation Kinetics and Rheological Properties. Polymers 9 (2017) 226.

[40] H. Yang, L.R. Shen, H. Bu, G. Li. Stable and biocompatible hydrogel composites based on collagen and dialdehyde carboxymethyl cellulose in a biphasic solvent system. Carbohyd. Polym. 222 (2019) 114974.

[41] X. Wang, O. Ronsin, B. Gravez, N. Farman, T. Baumberger, F. Jaisser, T. Coradin, C. Hélary. Nanostructured Dense Collagen-Polyester Composite Hydrogels as Amphiphilic Platforms for Drug Delivery. Adv. Sci. 8 (2021) 2004213.

[42] X. Zhao, X. Li, X. Xie, J. Lei, L. Ge, L. Yuan, D. Li, C. Mu. Controlling the Pore Structure of Collagen Sponge by Adjusting the Cross-Linking Degree for Construction of Heterogeneous Double-Layer Bone Barrier Membranes. ACS Appl. Bio Mater. 3 (2020) 2058−2067.

[43] H. Liang, S.J. Russell, D.J. Wood, G. Tronci. Monomer-Induced Customization of UV-Cured Atelocollagen Hydrogel Networks. Front. Chem. 6 (2018) 626.

[ 44 ] G. Tronci, J. Yin, R.A. Holmes, H. Liang, S.J. Russell, D.J. Wood. Protease-sensitive atelocollagen hydrogels promote healing in a diabetic wound model. J. Mater. Chem. B 4 (2016) 7249-7258.

[45] G. Tronci, S.J. Russell, D.J. Wood. Photo-active collagen systems with controlled triple helix architecture. J. Mater. Chem. B 1 (2013) 3705-3715.

[ 46 ] G. Tronci, C.A. Grant, N.H. Thomson, S.J. Russell, D.J. Wood. Multi-scale mechanical characterization of highly swollen photo-activated collagen hydrogels. J. R. Soc. Interface 12 (2015) 20141079.

[ 47 ] H. Liang, S.J. Russell, D.J. Wood, G. Tronci. A hydroxamic acid–methacrylated collagen conjugate for the modulation of inflammation-related MMP upregulation. J. Mater. Chem. B 6 (2018) 3703-3715.

[48] Z. Gao, X.B. Yang, E.J. Jones, P.A. Bingham, A. Scrimshire, P. Thornton, G. Tronci. An injectable, self-healing and MMP-inhibiting hyaluronic acid gel via iron coordination. Int. J. Biol. Macromol. 165 B (2020) 2022-2029.

[49] Y. Wang, R. Fu, X. Ma, X. Li, D. Fan. Development of a Mechanically Strong Nondegradable Protein Hydrogel with a Sponge-Like Morphology. Macromol Biosci 21 (2021) 2000396.

[50] A. Vajgel, N. Mardas, B. Carvalho Farias, A. Petrie, R. Cimões, N. Donos. A systematic review on the critical size defect model. Clin. Oral Implants Res. 25 (2014) 879-893.

[51] E. Calciolari, N. Mardas, X. Dereka, N. Kostomitsopoulos, A. Petrie, N. Donos. The effect of experimental osteoporosis on bone regeneration: Part 1, histology finding. Clin. Oral Implants Res. 28 (2017) e101-e110.

[52] J. Yin, D.J. Wood, S.J. Russell, G. Tronci. Hierarchically Assembled Type I Collagen Fibres as Biomimetic Building Blocks of Biomedical Membranes. Membranes 11 (2021) 620

[53] B. Brodsky, A.V. Persikov. Molecular Structure of the Collagen Triple Helix. Adv. Protein Chem. 70 (2005) 301-339.

[54] Z. Tian, W. Liu, G. Li. The microstructure and stability of collagen hydrogel crosslinked by glutaraldehyde. Polym. Degrad. Stab. 130 (2016) 264-270.

[55] M.M. Rad, S.N. Khorasani, L. Ghasemi-Mobarakeh, M.P. Prabhakaran, M. Reza Foroughi, M. Kharaziha, N. Saadatkish, S. Ramakrishna. Fabrication and characterization of two-layered nanofibrous membrane for guided bone and tissue regeneration application. Mater. Sci. Eng. C 80 (2017) 75-87.

[56] F. Yang, S.K. Both, X. Yang, X.F. Walboomers, J.A. Jansen. Development of an electrospun nano-apatite/PCL compositemembrane for GTR/GBR application. Acta Biomaterialia 5 (2009) 3295–3304.

[57] S.I. Somo, B. Akar, E.S. Bayrak, J.C. Larson, A.A. Appel, H. Mehdizadeh, A. Cinar, E.M. Brey. Pore Interconnectivity Influences Growth Factor-Mediated Vascularization in Sphere-Templated Hydrogels. Tissue Eng Part C Methods 21 (2015) 773-785.

[58] N. Vázquez-Portalatín, C.E. Kilmer, A. Panitch, J.C. Liu. Characterization of Collagen Type I and II Blended Hydrogels for Articular Cartilage Tissue Engineering. Biomacromolecules 17 (2016) 3145-3152.

[59] X. Zhang, L. Xu, S. Wei, M. Zhai, J. Li. Stimuli-responsive deswelling of radiation synthesized collagen hydrogel in simulated physiological environment. J. Biomed. Mater. Res. Part A 101 (2013) 2191-2201.

[60] W. B. Stockton, M. F. Rubner. Molecular-Level Processing of Conjugated Polymers. 4. Layer-by-Layer Manipulation of Polyaniline via Hydrogen-Bonding Interactions. Macromolecules 30 (1997) 2717-2725.

[61] B.S. Kim, J.S. Hrkach, R. Langer. Biodegradable photo-crosslinked poly(ether-ester) networks for lubricious coatings. Biomaterials 21 (2000) 259-265.

[62] A. Bozkurt, C. Apel, B. Sellhaus, S. van Neerven, B. Wessing, R.-D. Hilgers, N. Pallua. Differences in degradation behavior of two non-cross-linked collagen barrier membranes: an *in vitro* and in vivo study. Clin. Oral Impl. Res. 25 (2014) 1403-1411.

[63] B. Gurumurthy, J.A. Griggs, A.V. Janorkar. Optimization of collagen-elastin-like polypeptide composite tissue engineering scaffolds using response surface methodology. J. Mech. Behav. Biomed. Mater. 2018 (84) 116-125.

[64] N.A. Peppas, J.Z. Hilt, A. Khademhosseini, R. Langer. Hydrogels in Biology and Medicine: From Molecular Principles to Bionanotechnology. Adv. Mater. 18 (2006) 1345-1360.

[65] S.V. Kanth, A. Ramaraj, J.R. Rao, B.U. Nair. Stabilization of type I collagen using dialdehyde cellulose. Process Biochem. 44 (2009) 869-874.

[66] M. Wang, J. Li, W. Li, Z. Du, S. Qin. Preparation and characterization of novel poly(vinyl alcohol)/collagen double-network hydrogels. Int. J. Biol. Macromol. 118 (2018) 41-48.

[67] C. Mu, K. Zhang, W. Lin, D. Li. Ring-opening polymerization of genipin and its long-range crosslinking effect on collagen hydrogel. J. Biomed. Mater. Res. Part A 101 (2013) 385-393.

[68] M. Kahl, D. Schneidereit, N. Bock, O. Friedrich, D.W. Hutmacher, C. Meinert. *MechAnalyze*: An Algorithm for Standardization and Automation of Compression Test Analysis. Tissue Eng. Part C Methods 27 (2021) 529-542.

[69] G.M. Fernandes‑Cunha, K.M. Chen, F. Chen, P. Le, J.H. Han, L.A. Mahajan, H.J. Lee, K.S. Na, D. Myung. In situ‑forming collagen hydrogel crosslinked via multi‑functional PEG as a matrix therapy for corneal defects. Sci. Rep. 10 (2020) 16671.

[70] P. Bhaskar, L.A. Bosworth, R. Wong, M.A. O'brien, H. Kriel, E. Smit, D.A. McGrouther, J.K. Wong, S.H. Cartmell. Cell response to sterilized electrospun poly($\varepsilon$-caprolactone) scaffolds to aid tendon regeneration *in vivo*. J Biomed Mater Res Part A 105A (2017) 389-397.

[71] R. Ghobeira, C. Philips, H. Declercq, P. Cools, N. De Geyter, R. Cornelissen, R. Morent. Effects of different sterilization methods on the physico-chemical and bioresponsive properties of plasma-treated polycaprolactone films. Biomed. Mater. 12 (2017) 015017.

[72] C. Somaiah, A. Kumar, D. Mawrie, A. Sharma, S.D. Patil, J. Bhattacharyya, R. Swaminathan, B.G. Jaganathan. Collagen Promotes Higher Adhesion, Survival and Proliferation of Mesenchymal Stem Cells. PLoS ONE 10 (2015) e0145068.

[73] S. Piluso, A. Lendlein, A.T. Neffe. Enzymatic action as switch of bulk to surface degradation of clicked gelatin-based networks. Polym. Adv. Technol. 28 (2017) 1318-1324.

[74] N. Donos, X. Dereka, N. Mardas. Experimental models for guided bone regeneration in healthy and medically compromised conditions. Periodontol. 2000 68 (2015) 99-121.

[75] N. Donos, N.P. Lang, I.K. Karoussis, D. Bosshardt, M. Tonetti, L. Kostopoulos. Effect of GBR in combination with deproteinized bovine bone mineral and/or enamel matrix proteins on the healing of critical-size defects. Clin. Oral Implants Res. 15 (2004) 101-111.

[76] D. Lundgren, S. Nyman, T. Mathisen, S. Isaksson, B. Klinge. Guided bone regeneration of cranial defects, using biodegradable barriers: An experimental pilot study in the rabbit. J. Craniomaxillofac. Surg. 20 (1992) 257-260.

[77] E. Calciolari, N. Mardas, X. Dereka, N. Kostomitsopoulos, A. Petrie, N. Donos. The effect of experimental osteoporosis on bone regeneration: Part 1, histology findings. Clin. Oral Implants Res. 28 (2017) e101-e110.

[78] S. Saha, X.B. Yang, N. Wijayathunga, S. Harris, G.A. Feichtinger, R.P.W. Davies, J. Kirkham. A biomimetic self-assembling peptide promotes bone regeneration *in vivo*: A rat cranial defect study. Bone 127 (2019) 602-611.

[79] R. Al-Kattan, M. Retzepi, E. Calciolari, N. Donos. Microarray gene expression during early healing of GBR-treated calvarial critical size defects. Clin. Oral Implants Res. 28 (2017) 1248-1257.

[80] N. Dubey, J.A.Ferreira, A. Daghrery, Z. Aytac, J. Malda, S.B. Bhaduri, M.C. Bottino. Highly tunable bioactive fiber-reinforced hydrogel for guided bone regeneration. Acta Biomaterialia 113 (2020) 164-176.

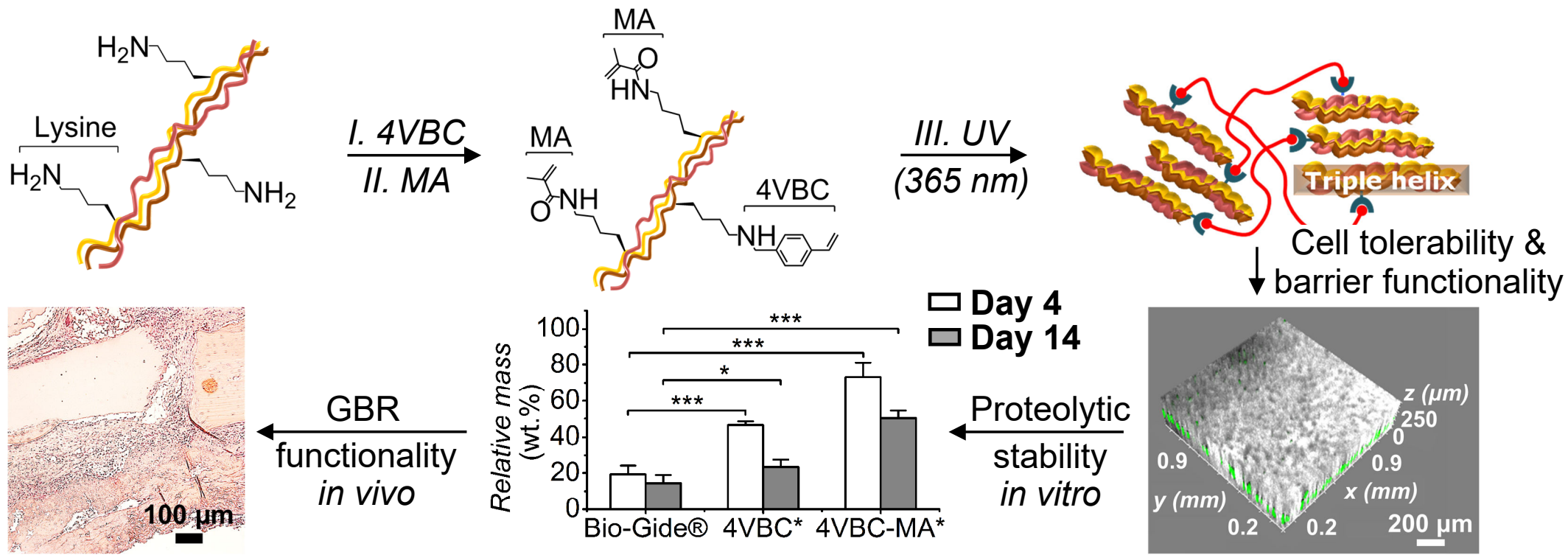

**Figure captions**

**Figure 1.** Design of an AC membrane with increased proteolytic stability. (I-II): AC is sequentially functionalised with both 4-vinylbenzyl chloride (4VBC, I) and methacrylic anhydride (MA, II). (III): The reaction product is solubilised in an aqueous solution supplemented with a water-soluble photoinitiator and UV-cured, to generate a covalently crosslinked network of sequentially functionalised AC triple helices.

**Figure 2.** Far-UV CD spectra of native (light grey) and sequentially functionalised (black) AC, indicating the presence of polyproline-II (negative peak at 197 nm) and triple (positive peak at 221 nm) helices.

**Figure 3.** Cool-stage electron microscopy of a typical sample of hydrogel 4VBC-MA* prepared in the I2959-supplemented PBS solution. (A-B): top surface at low (A) and high (B) magnification; (C) cross-section.

**Figure 4.** Exemplary profiles of storage ($G'$, black line) and loss ($G''$, light grey line) modulus recorded via oscillatory time sweeps under UV light. Samples of 4VBC-MA were dissolved in I2959-supplemented solutions of either 10 mM HCl (A), 17.4 mM AcOH (B) or 10 mM PBS (C) and UV light-activated 5 s following shear oscillation.

**Figure 5.** Swelling ratio ($SR$, grey column, left hand side Y axis) and gel content ($G$, white column, right hand side Y axis) of samples 4VBC-MA* UV-cured in either 10 mM HCl, 17.4 mM AcOH or 10 mM PBS. $SR$ and $G$ were measured in distilled water and 17.4 mM AcOH, respectively. Data are presented as mean ± SD (n=4). *$p < 0.05$, **$p < 0.01$ and ***$p < 0.001$. (B) Exemplary macrographs of sample 4VBC-MA* in both dry and fully hydrated state.

**Figure 6.** (A): stress-compression curves (A) of the water-equilibrated AC networks UV-cured in either 10 mM HCl (grey), 17.4 mM AcOH (blue) or 10 mM PBS (black). (B): Compression modulus calculated from previous compression curves ($\varepsilon$= 25-30%). Data are presented as mean ± SD (n=3), ***$p < 0.005$.

**Figure 7.** Stability test *in vitro* and *in vivo*. (A-B): Relative mass ($\mu_{rel}$) of samples 4VBC*, 4VBC-MA* and Bio-Gide® at day 4 (white) and 14 (grey) of enzymatic incubation *in vitro* (37 °C, 10 mM PBS). (A): 1 CDU·ml$^{-1}$; (B): 2.5 CDU·ml$^{-1}$. Data (n=4) are presented as mean ± standard deviation. *$p< 0.05$, ***$p< 0.005$. (C): Digital macrographs of samples 4VBC-MA* and Bio-Gide® during enzymatic incubation (37 °C, 10 mM PBS, 2.5 CDU·ml$^{-1}$). (D): Stability of hydrogel 4VBC-MA* following 4-week subcutaneous implantation in rats. Scale bar ≈ 5 mm.

**Figure 8.** Confocal microscopy of live/dead L929 fibroblasts following 7-day culture on sample 4VBC-MA*. (A): cell-free hydrogel; (B): live (green) cells; (C): dead (red) cells; (D): superimposition of (A)-(C). Scale bar ~ 200 μm.

**Figure 9.** Cell viability and growth on either hydrogel 4VBC-MA* or tissue culture plastic

during 7 days of culture. (A): Relative metabolic activity of L929 cells cultured on either hydrogel 4VBC-MA* (black) or tissue culture plastic (grey) with respect to day 1. Data are determined via Alamar Blue assay, and presented relative to day 1 as mean ± standard deviation. (B): Live/dead staining of L929 fibroblasts after 7-day culture on hydrogel 4VBC-MA*. (C-D): Cool-stage SEM images of hydrogel 4VBC-MA* at day 7 of L929 cell culture.

**Figure 10.** *In vivo* assessment of 4VBC-MA* and Bio-Gide® in a GBR model. (A): Graphical representation of the GBR calvarial defect model (Ø 5 mm) and implantation of membrane samples. The intracranial side (i.s.) is always covered with Bio-Gide®, whilst the extracranial side (e.s.) is covered by either the hydrogel 4VBC-MA* or Bio-Gide®. (B-I): H&E staining of *ex vivo* sections collected following 1-week implantation. Black arrows point to the bone filler (Bio-Oss®). Sample codes are indicated in the images according to Table 1; *w*: week; *B*: Bio-Gide®; *C*: 4VBC-MA* (Collagen); *BO*: Bio-Oss®. Scale bars: ~ 400 µm (B,D,F,H) and ~ 100 µm (C,E,G,I).

**Figure 11.** H&E staining of *ex vivo* sections collected following 4-week implantation of 4VBC-MA* and Bio-Gide® in the GBR model in rats. Black arrows point to the bone filler (Bio-Oss®), blue arrows indicate densely packed tissue formation. Sample codes are indicated in the images according to Table 1; *w*: week; *B*: Bio-Gide®; *C*: 4VBC-MA* (Collagen); *BO*: Bio-Oss®. Scale bars: ~ 400 µm (A,C,E,G) and ~ 100 µm (B,D,F,H).

Figure 1

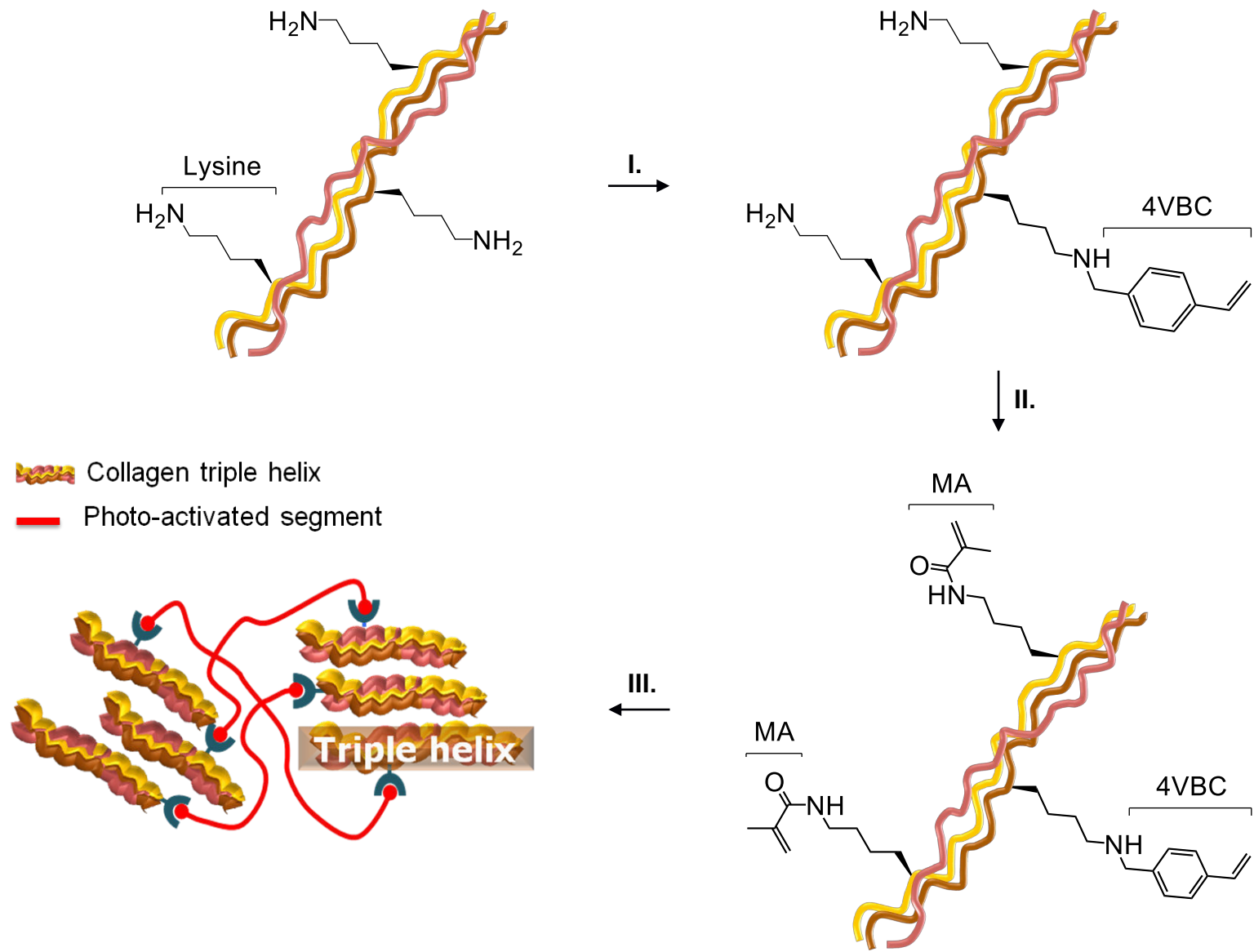

Figure 2

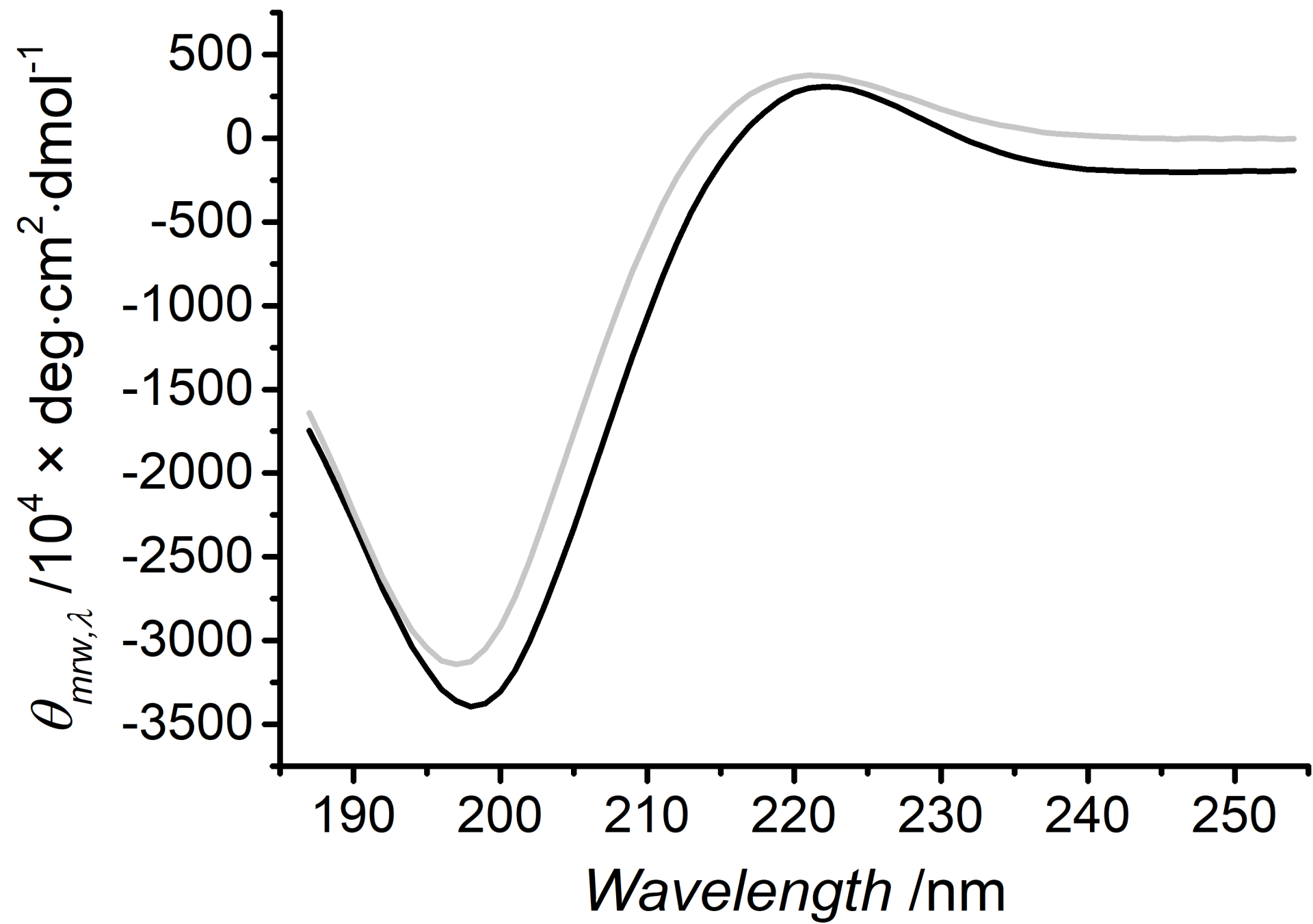

Figure 3	Click here to access/download;Figure(s);Figure 3.pdf

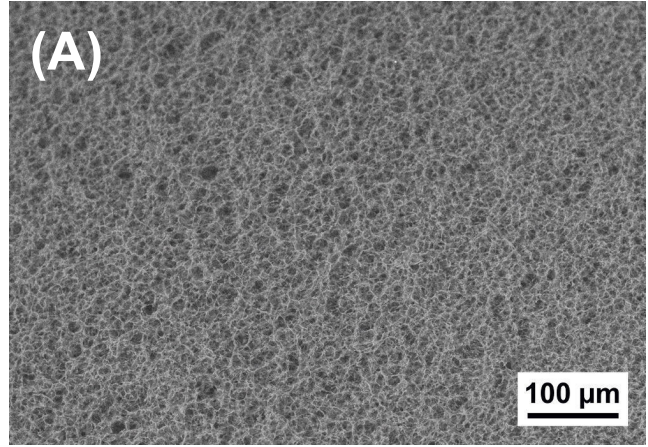 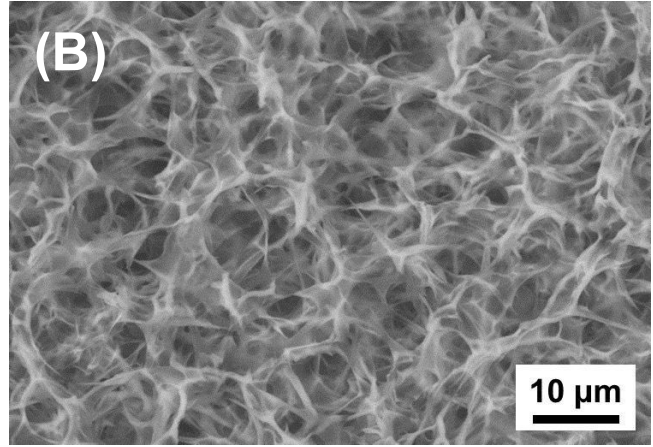 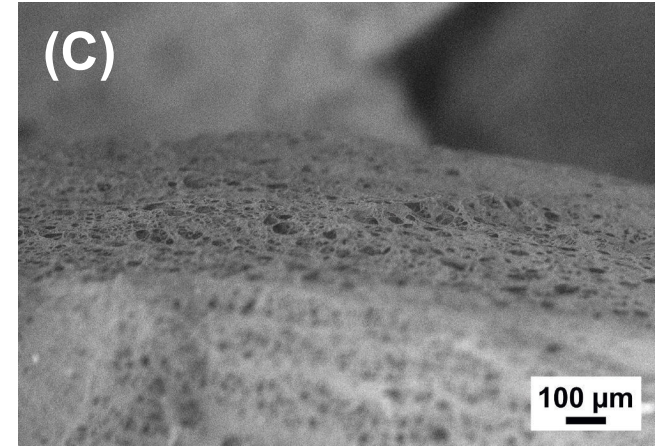

Figure 4

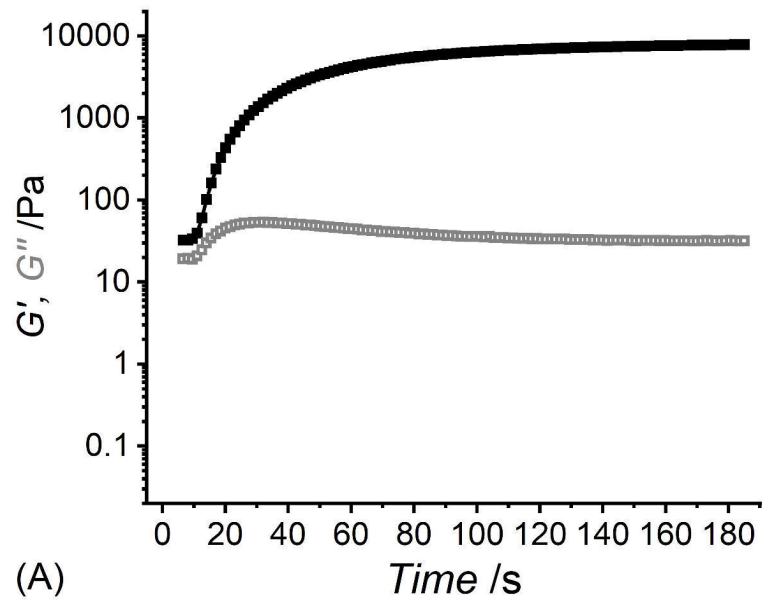 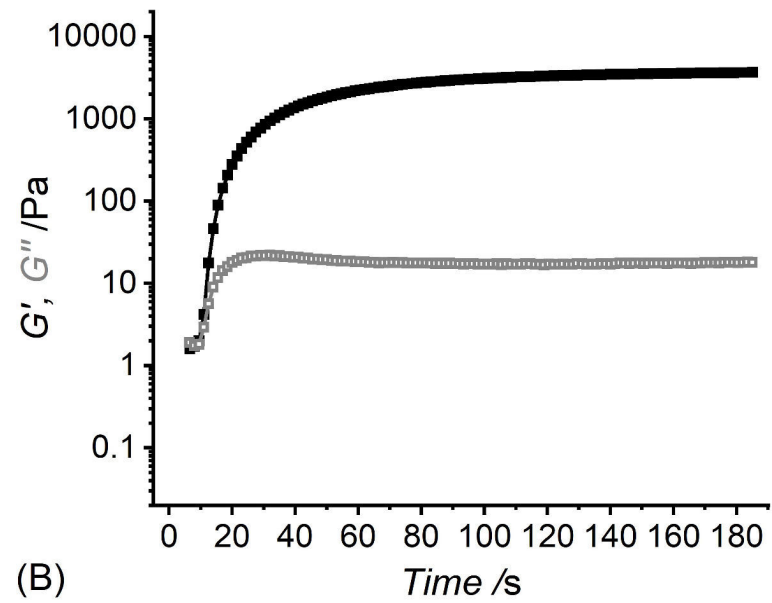 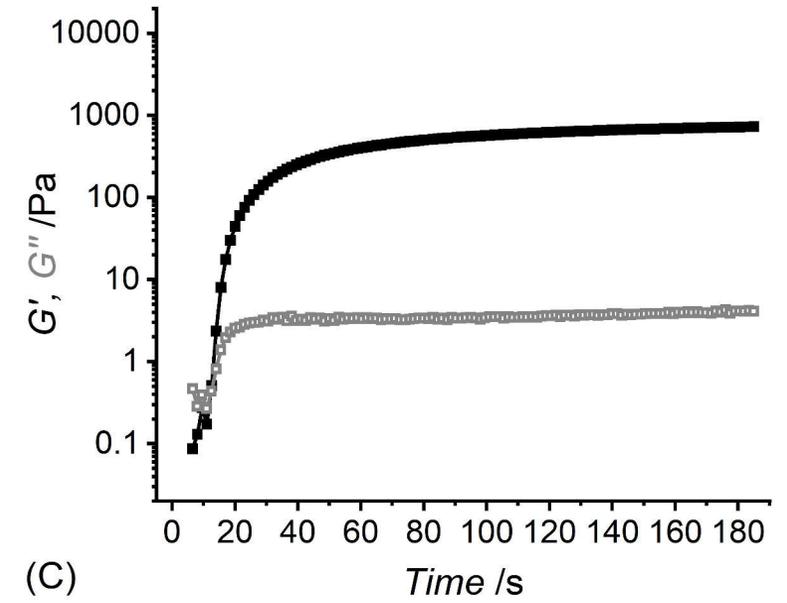

(A) (B) (C)

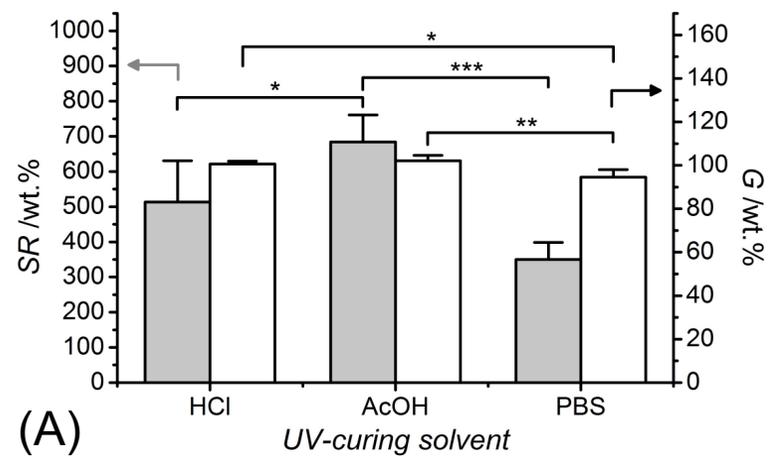
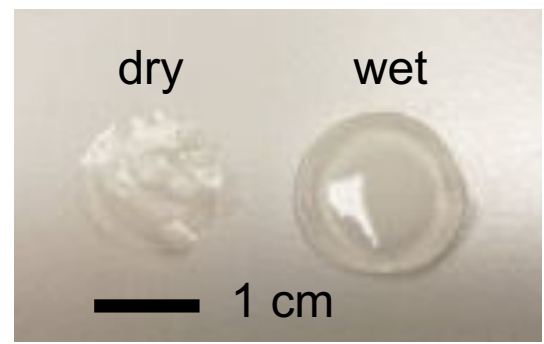

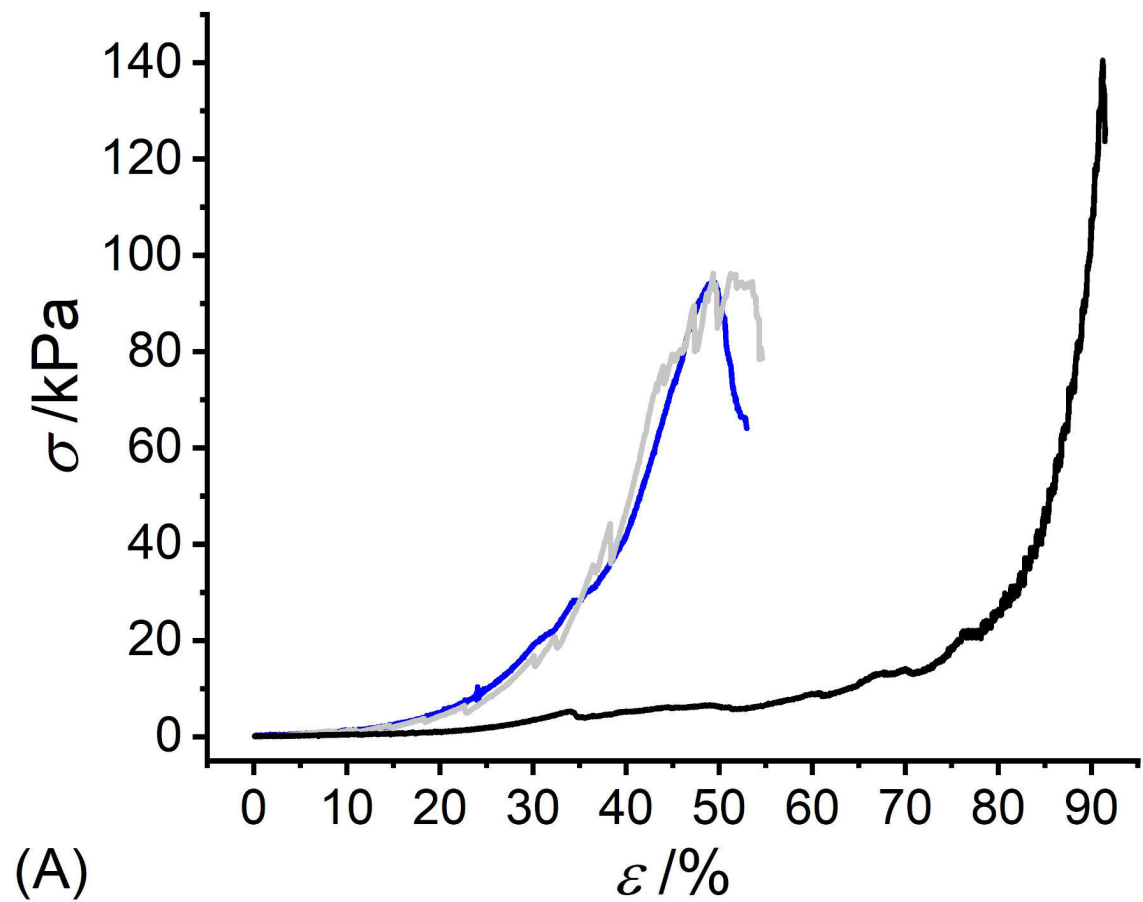
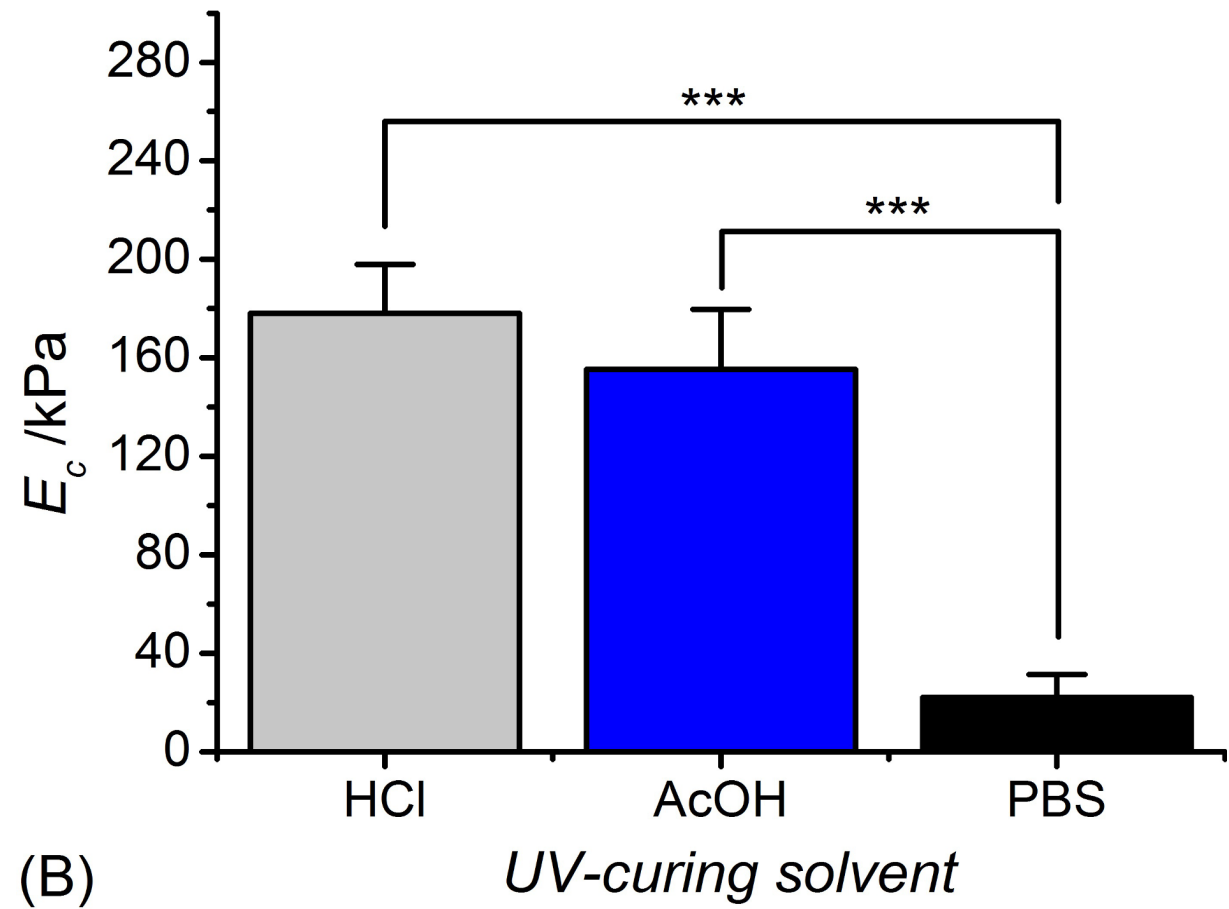

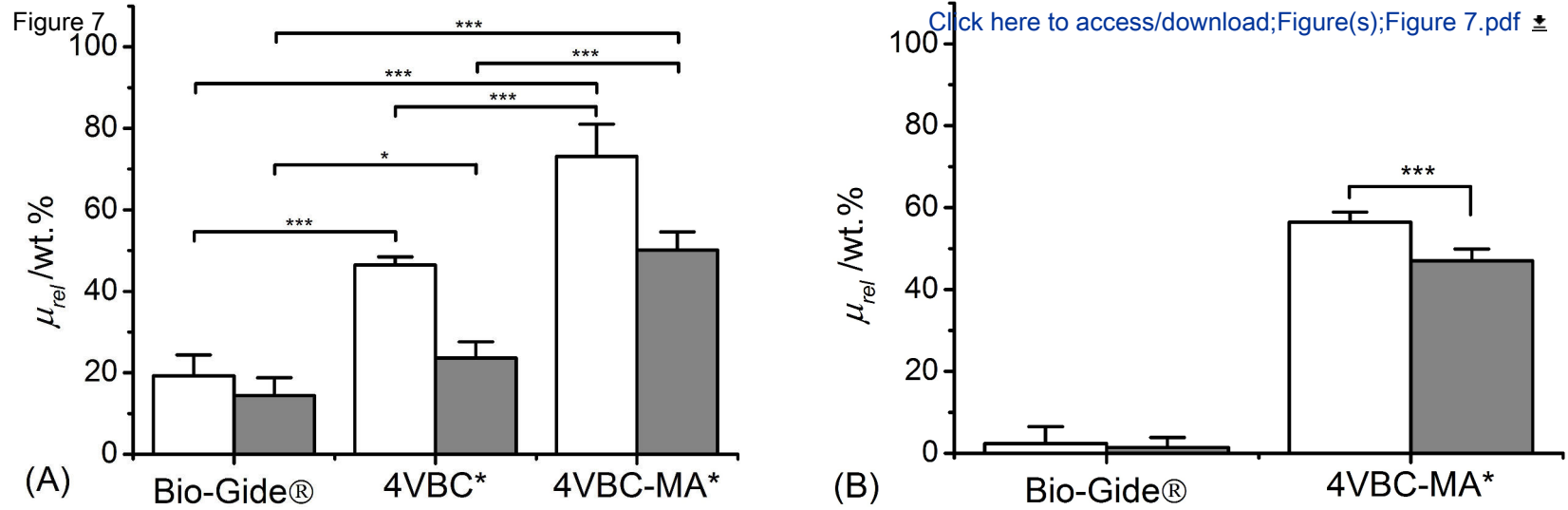
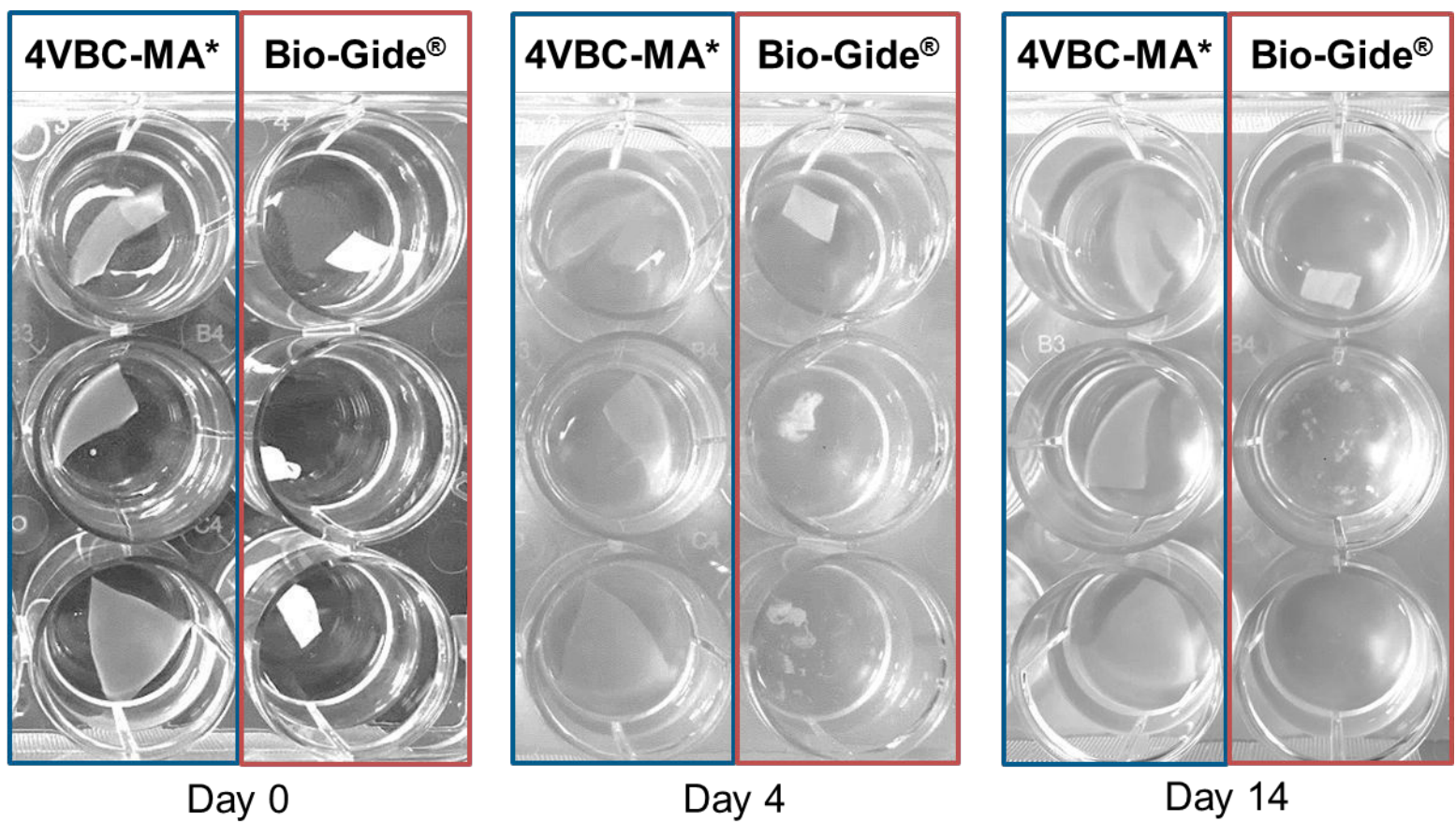
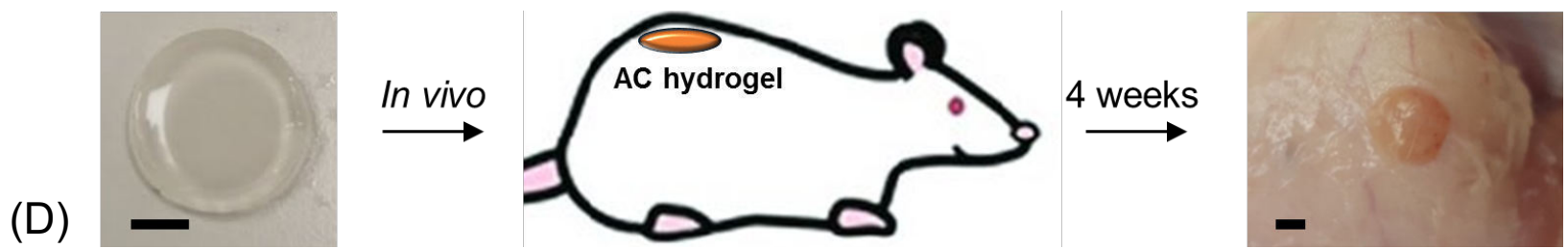

Figure 7

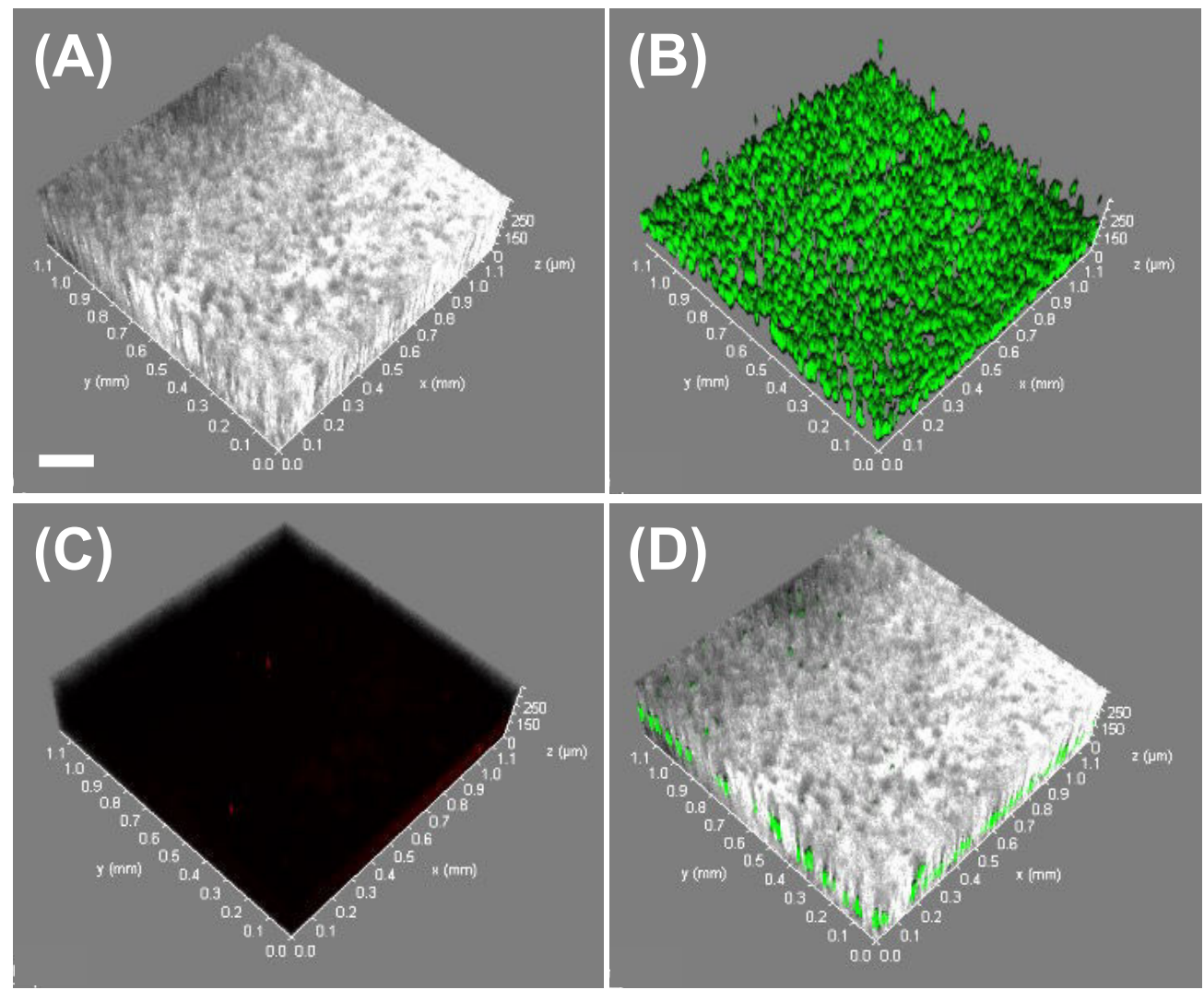

Figure 9

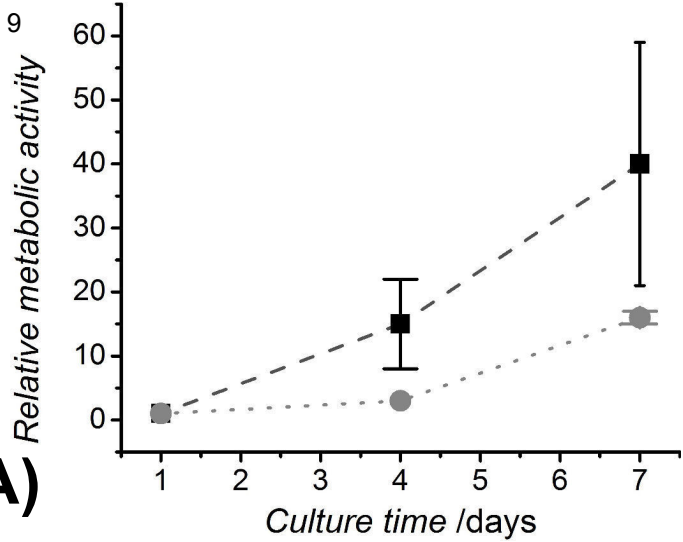
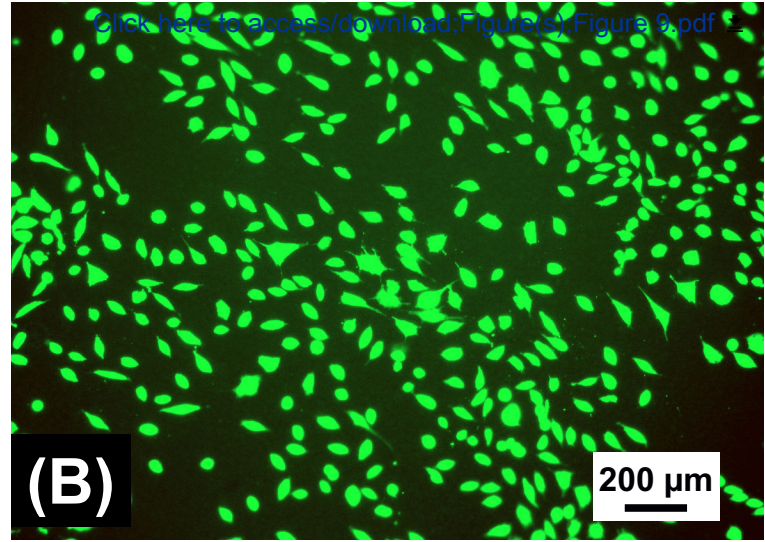
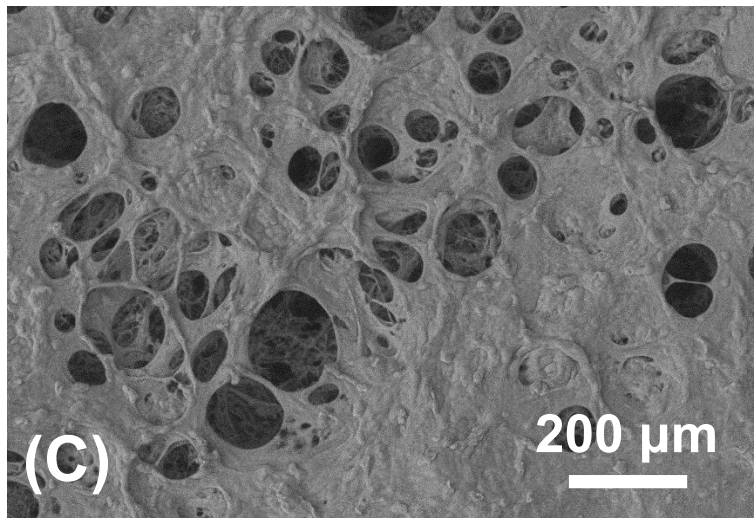
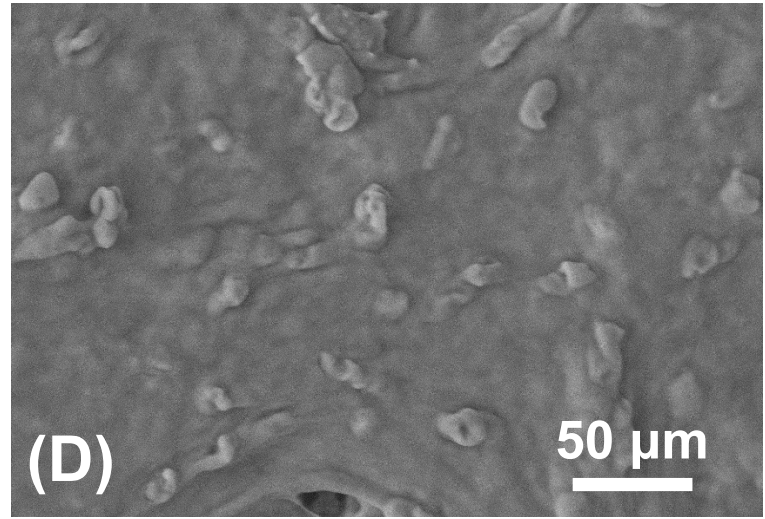

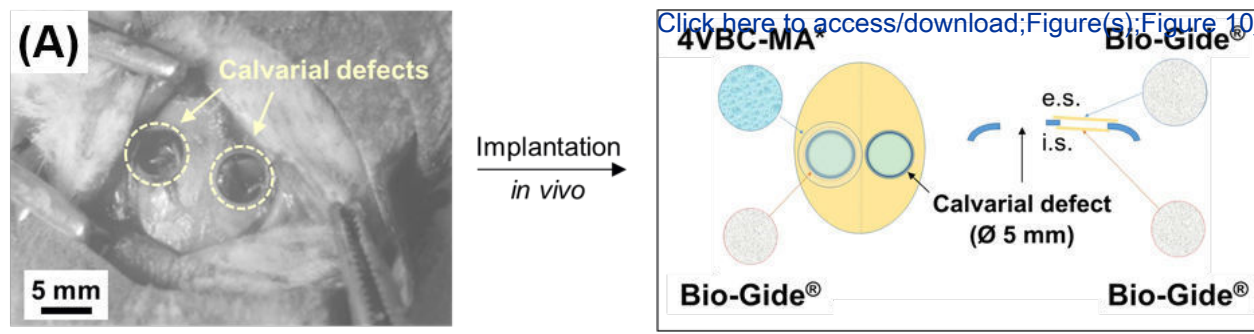
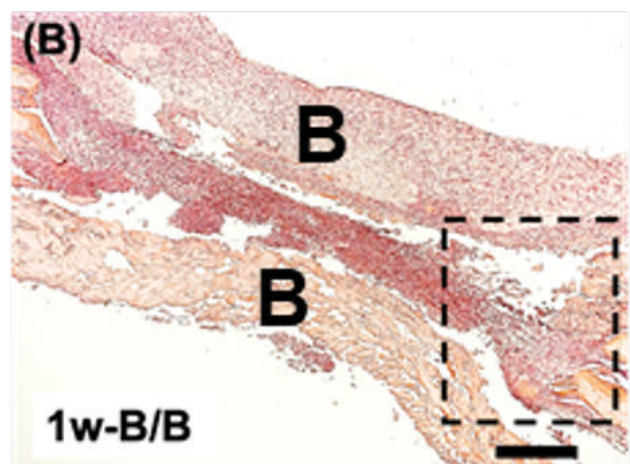
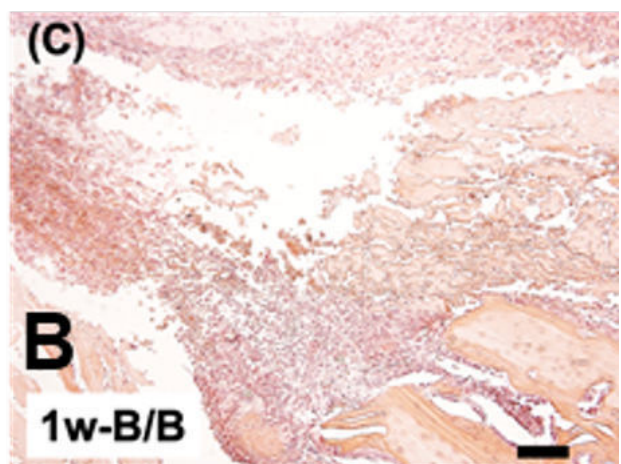
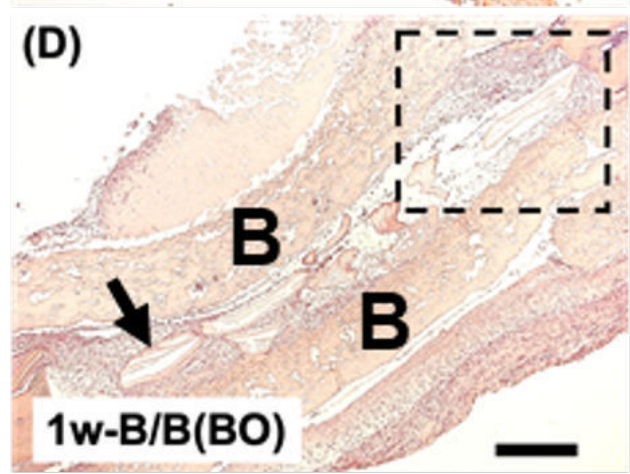
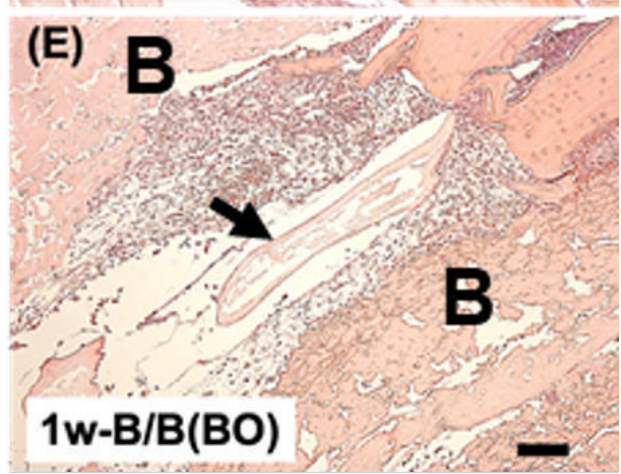
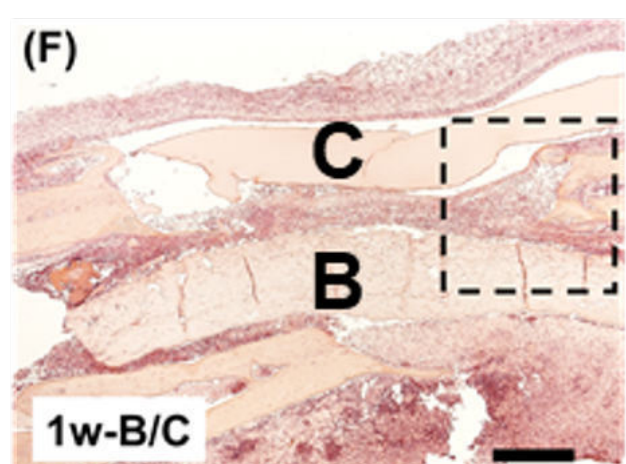
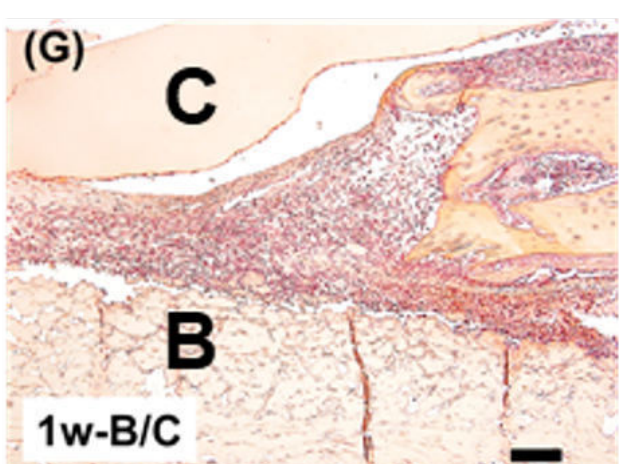
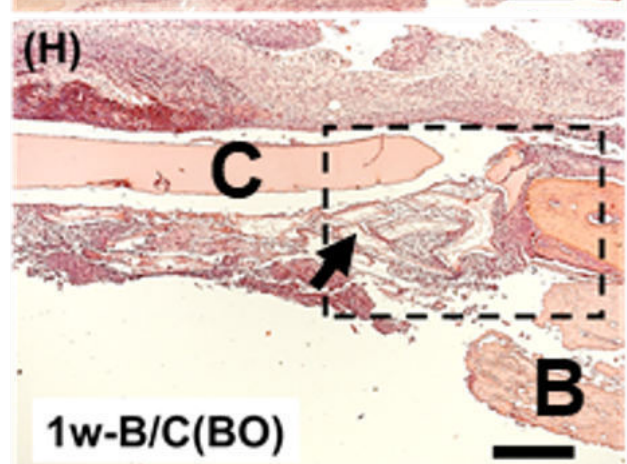
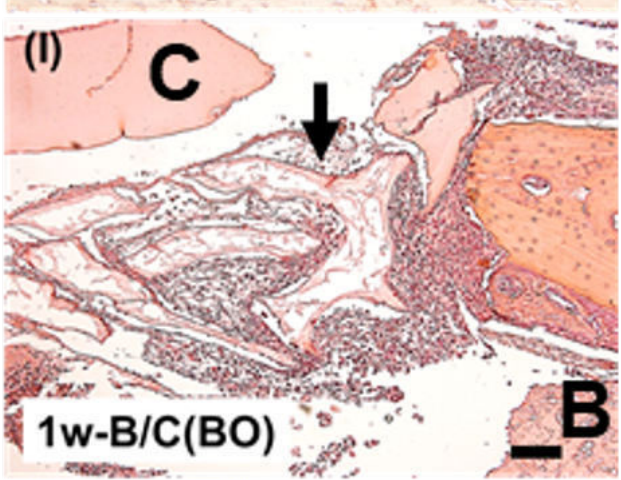

Figure 11

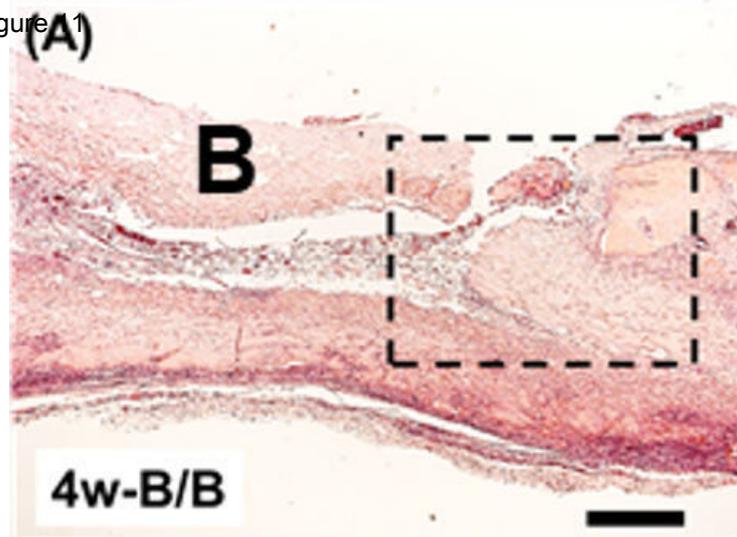
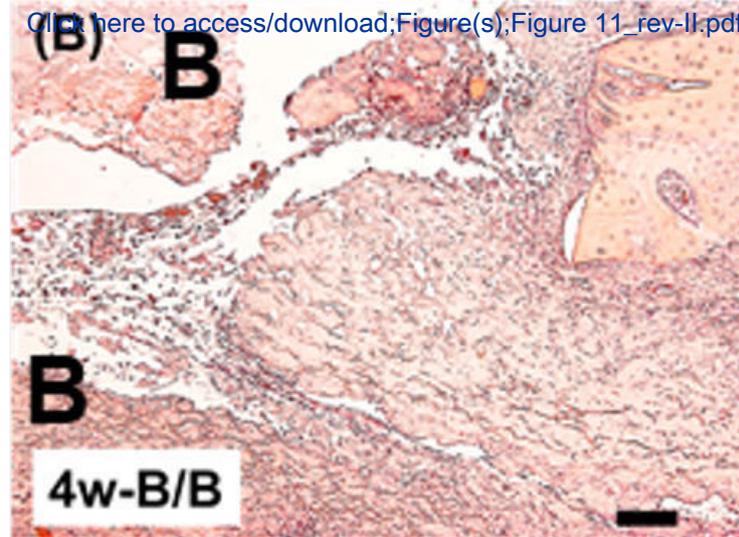
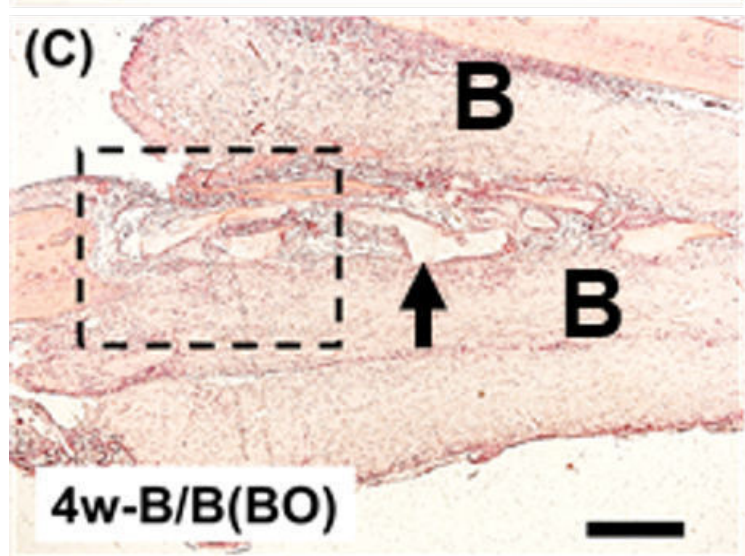
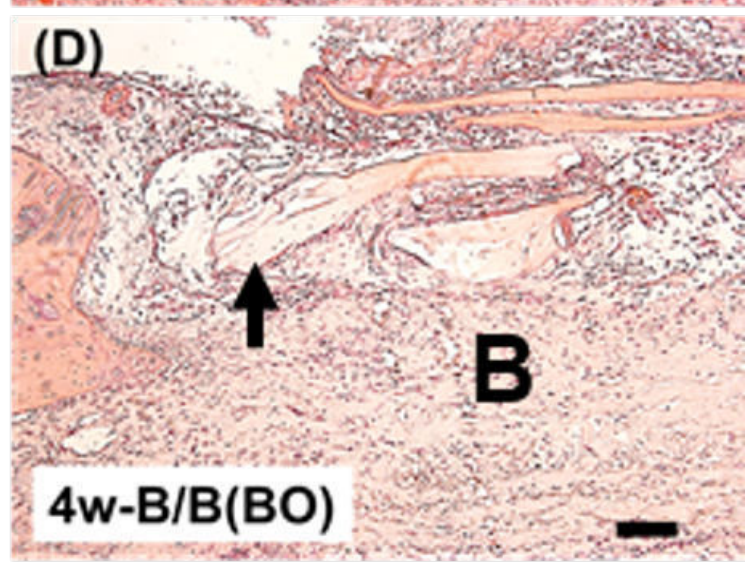
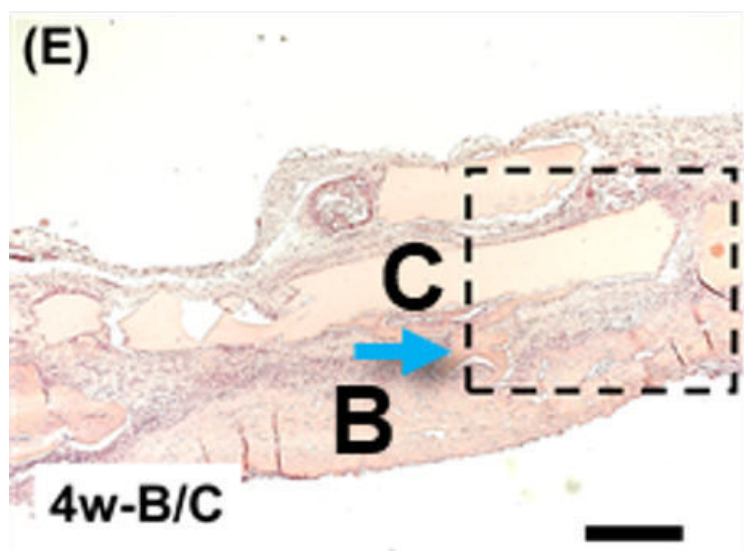
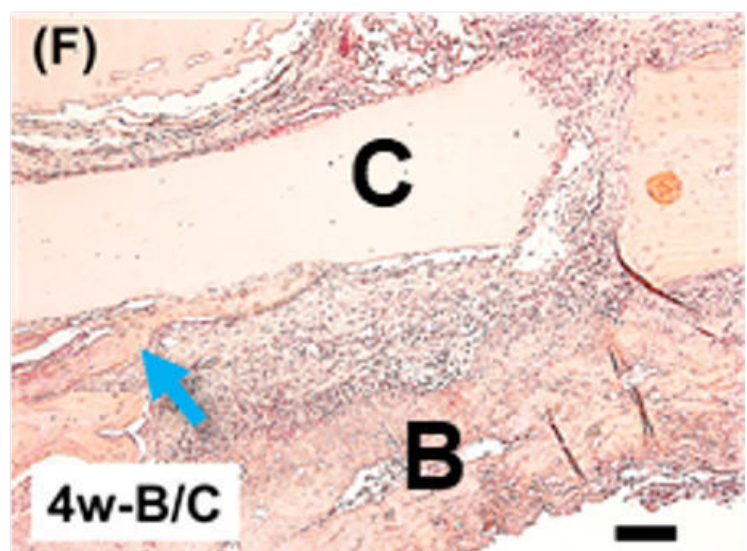
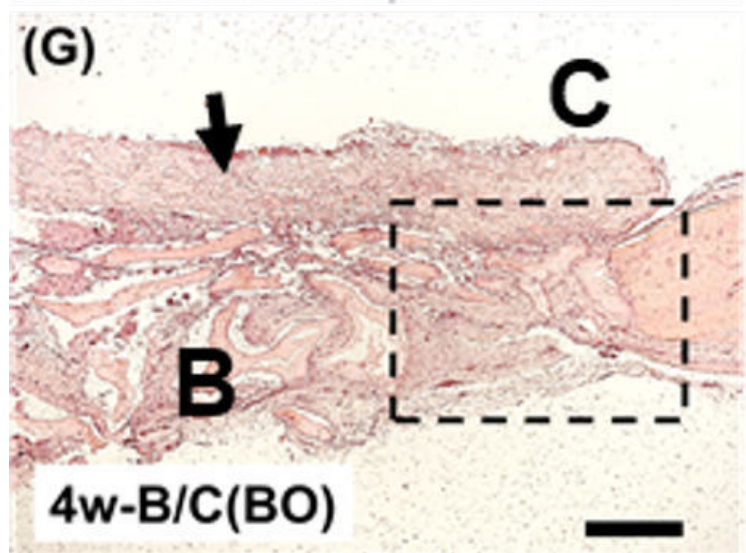
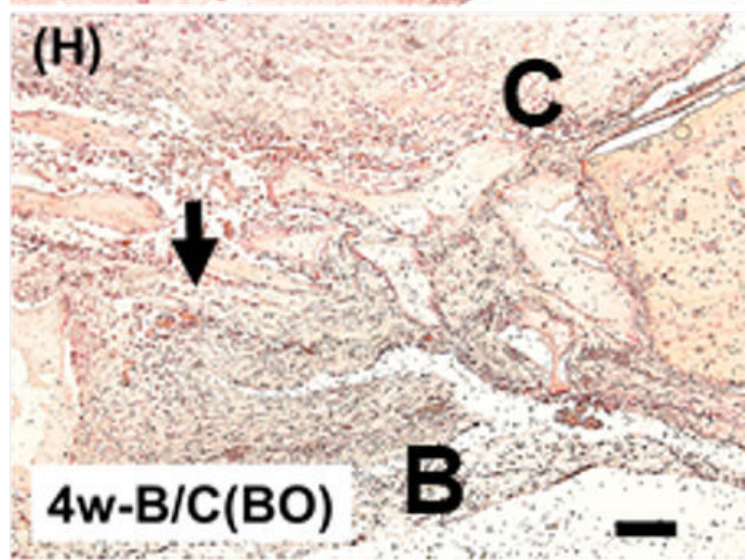

**Tables**

**Table 1.** Experimental groups collected after 1- and 4-week implantation *in vivo*. Sample ID: *w*: week; *B*: Bio-Gide®; *C*: 4VBC-MA*; *BO*: Bio-Oss®. Two replicates were used for each group. n.a.: not applicable.

| Group | Sample ID | Bottom | Top | Filler |
|---|---|---|---|---|
| 1 | 1w-BB | Bio-Gide® | Bio-Gide® | n.a. |
| 2 | 1w-BB-BO | Bio-Gide® | Bio-Gide® | Bio-Oss® |
| 3 | 1w-BC | Bio-Gide® | 4VBC-MA* | n.a. |
| 4 | 1w-BC-BO | Bio-Gide® | 4VBC-MA* | Bio-Oss® |
| 5 | 4w-BB | Bio-Gide® | Bio-Gide® | n.a. |
| 6 | 4w-BB-BO | Bio-Gide® | Bio-Gide® | Bio-Oss® |
| 7 | 4w-BC | Bio-Gide® | 4VBC-MA* | n.a. |
| 8 | 4w-BC-BO | Bio-Gide® | 4VBC-MA* | Bio-Oss® |

**Table 2.** Degree of functionalisation (*F*) of 4VBC- and sequentially functionalised AC samples (n=3) reacted with defined monomer-to-lysine molar ratios ($[M] \cdot [Lys]^{-1}$). Both TNBS and Ninhydrin assays were employed for the quantification of remaining lysine amine groups. n.a.: not applicable. n.o.: not observed.

| Sample ID | $[M] \cdot [Lys]^{-1}$ | [Lys] /mol·g$^{-1}$ (×10$^{-4}$) | | *F* /mol.% |
|---|---|---|---|---|
| | | TNBS | Ninhydrin | |
| AC | n.a. | 2.89 ± 0.06 | n.o. | n.a. |
| 4VBC | 25 | 2.38 ± 0.02 | 2.38 ± 0.01 | 18 ± 1 |
| 4VBC-MA | 25[a] | 0.1 ± 0.06 | 0.17 ± 0.05 | 97 ± 2 |

[a] Sample obtained via subsequent reaction of sample 4VBC with MA ($[MA] \cdot [Lys]^{-1}$ = 25).

## Supporting Information

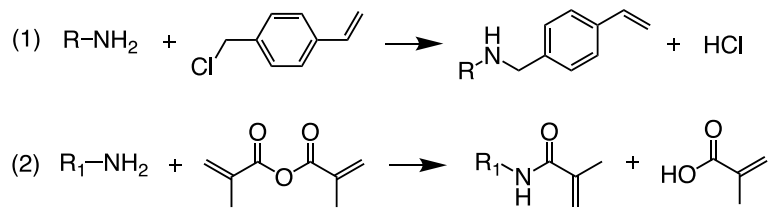

**Figure S1.** Reaction of atelocollagen with 4VBC (1) and MA (2), yielding photoactive products and hydrochloric acid and methacrylic acid by-products, respectively. R: AC; $R_1$: 4VBC-functionalised AC.

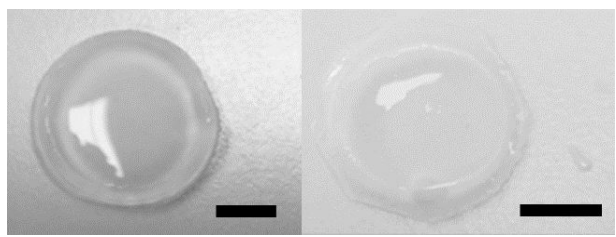

**Figure S2.** Macroscopic images of UV-cured samples 4VBC-MA* (left) and 4VBC* (right). Scale bar ≈ 5 mm. Samples of 4VBC-MA* and 4VBC* were prepared in an I2959-supplemented solution of 10 mM PBS and 10 mM HCl, respectively.

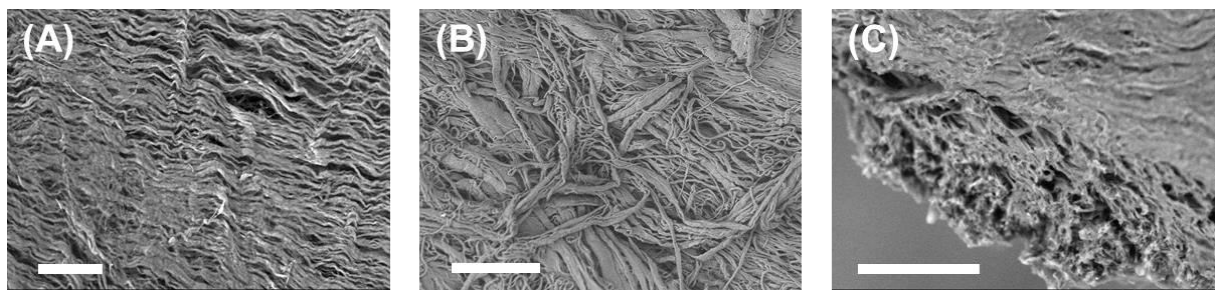

**Figure S3.** Scanning electron microscopy of commercial collagen membrane (Bio-Gide®, Geistlich). (A): bottom surface; (B): top surface; (C): cross-section. Scale bar ~ 100 μm.

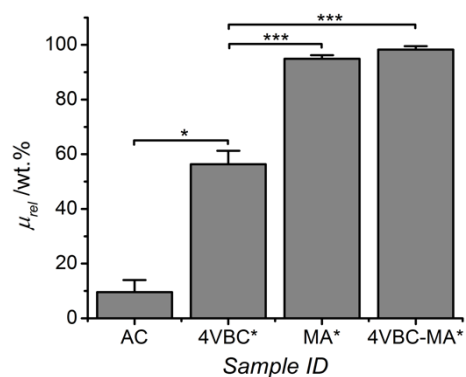

**Figure S4.** Relative mass ($\mu_{rel}$, n=4) of AC and UV-cured samples 4VBC*, MA* and 4VBC-MA* following 4-day incubation in collagenase-supplemented media (37 °C, 10 mM PBS, 5 CDU·ml$^{-1}$. Samples of 4VBC*, MA* and 4VBC-MA* were cured in 10 mM HCl solution supplemented with I2959. * $p< 0.05$; ** $p< 0.01$; *** $p< 0.005$.

**Table S1.** Storage (*G'*) and shear (*G''*) moduli of hydrogel 4VBC-MA* (n=3) obtained following UV-curing of the sequentially functionalised atelocollagen precursor in varied aqueous environments. Data in brackets refer to the shear moduli of the hydrogel-forming solution prior to UV curing.

| *UV curing solvent* | *G' /Pa* | *G'' /Pa* |
|---|---|---|
| 10 mM HCl (pH 2) | 8149 ± 349 (39) | 40 ± 12 (21) |
| 17.4 mM AcOH (pH 3) | 3675 ± 185 (4) | 19 ± 1 (3) |
| 10 mM PBS (pH 7.5) | 796 ± 61 (0.2) | 5 ± 1 (0.3) |